\documentclass[conference]{IEEEtran}
\IEEEoverridecommandlockouts
\usepackage{cite}
\usepackage{amsmath,amssymb,amsfonts}
\usepackage{algorithmic}
\usepackage{algorithm}
\usepackage{graphicx}
\usepackage{textcomp}
\usepackage{xcolor}
\usepackage{url}
\usepackage{caption}
\usepackage{subcaption}
\usepackage{booktabs}
\usepackage{enumitem}
\usepackage{tikz}
\usepackage{pgfplots}
\pgfplotsset{compat=1.18}

\def\BibTeX{{\rm B\kern-.05em{\sc i\kern-.025em b}\kern-.08em
    T\kern-.1667em\lower.7ex\hbox{E}\kern-.125emX}}
\begin{document}

\makeatletter
\newcommand{\linebreakand}{%
  \end{@IEEEauthorhalign}
  \hfill\mbox{}\par
  \mbox{}\hfill\begin{@IEEEauthorhalign}
}
\makeatother

\title{Enhancing HNSW Index for Real-Time Updates: Addressing Unreachable Points and Performance Degradation
}

\author{
\IEEEauthorblockN{1\textsuperscript{st} Wentao Xiao}
\IEEEauthorblockA{\textit{SCSE} \\
\textit{UESTC}\\
Chengdu, China \\
WentaoXiao1234@gmail.com}
\and
\IEEEauthorblockN{2\textsuperscript{nd} Yueyang Zhan}
\IEEEauthorblockA{\textit{SCSE} \\
\textit{UESTC}\\
Chengdu, China \\
yueyang.zhan@std.uestc.edu.cn
}
\and
\IEEEauthorblockN{3\textsuperscript{rd} Rui Xi*}
\IEEEauthorblockA{\textit{SCSE} \\
\textit{UESTC}\\
Chengdu, China \\
ruix.ryan@gmail.com
{\thanks{*Corresponding author}}
}
\linebreakand
\IEEEauthorblockN{4\textsuperscript{th} Mengshu Hou}
\IEEEauthorblockA{\textit{SCSE} \\
\textit{UESTC}\\
Chengdu, China \\
mshou@uestc.edu.cn}
\and
\IEEEauthorblockN{5\textsuperscript{th} Jianming Liao}
\IEEEauthorblockA{\textit{SCSE} \\
\textit{UESTC}\\
Chengdu, China \\
liaojm@uestc.edu.cn}
}

\maketitle

\begin{abstract}
The approximate nearest neighbor search (ANNS) is a fundamental and essential component in data mining and information retrieval, with graph-based methodologies demonstrating superior performance compared to alternative approaches. Extensive research efforts have been dedicated to improving search efficiency by developing various graph-based indices, such as HNSW (Hierarchical Navigable Small World). However, the performance of HNSW and most graph-based indices become unacceptable when faced with a large number of real-time deletions, insertions, and updates. Furthermore, during update operations, HNSW can result in some data points becoming unreachable, a situation we refer to as the `unreachable points phenomenon'. This phenomenon could significantly affect the search accuracy of the graph in certain situations.

To address these issues, we present efficient measures to overcome the shortcomings of HNSW, specifically addressing poor performance over long periods of delete and update operations and resolving the issues caused by the unreachable points phenomenon. Our proposed MN-RU algorithm effectively improves update efficiency and suppresses the growth rate of unreachable points, ensuring better overall performance and maintaining the integrity of the graph. Our results demonstrate that our methods outperform existing approaches. Furthermore, since our methods are based on HNSW, they can be easily integrated with existing indices widely used in the industrial field, making them practical for future real-world applications. Code is available at \url{https://github.com/xwt1/MN-RU.git}
\end{abstract}


\section{Introduction}
The approximate nearest neighbor search (ANNS) has emerged as a focal point in diverse application areas encompassing data mining, information retrieval, and recommendation systems~(\cite{10.14778/3476249.3476255,10.1145/3511808.3557098}). Notably, deep learning models, such as large language models (LLMs), possess the capability to encode various data types, ranging from textual documents to visual and auditory inputs, into vector representations. Innovative systems like Retrieval-Augmented Generation (RAG)~(\cite{10.5555/3495724.3496517}) leverage ANNS to efficiently retrieve relevant documents or information, integrating them with generative model outputs to augment the precision and relevance of the generated content.
The basic ANNS can be defined as follows: Given a data set \(D \subseteq \mathbb{R}^{d}\) with \(n\) points in some metric space, a query data point \(q \in \mathbb{R}^{d}\), and \(k \in \mathbb{N}\), we seek to find the set \(L\) that efficiently represents the \(k\) nearest neighbors of \(q\) from \(P\), while maximizing \(\text{recall@}k = \frac{|G \cap L|}{|G|}\), where \(|G|=|L|=k\). Here, \(G\) is the ground truth set of \(q\)'s \(k\) nearest neighbors in \(P\). Effective indexing is crucial in this context, as it significantly enhances the efficiency and accuracy of the search process, enabling the rapid and reliable retrieval of nearest neighbors from large datasets.

Based on the theory and concepts of ANNS, a large number of indexing methods have been designed and can generally be classified into four types: tree structure-based index \cite{Muja2014ScalableNN}, hashing-based index \cite{jiang2015scalable}, quantization-based index \cite{johnson2019billion}, and graph-based index~(\cite{10.14778/3303753.3303754,10.14778/3489496.3489506,10.1145/3543507.3583552,10.1145/3639269}). Among these, graph-based methods demonstrate superior empirical search performance compared to others~(\cite{AUMULLER2020101374}).

Nevertheless, despite the commendable search performance exhibited by certain early graph-based techniques, a notable limitation lies in their static nature~(\cite{10.14778/3303753.3303754,10.14778/3489496.3489506}), rendering them incapable of accommodating the real-time modifications requisite in practical applications. For instance, consider the RAG system, which relies on kNN-based methodologies to transform documents into vector representations. In scenarios where users seek to alter the content within the system, such modifications entail the insertion of new data points and the removal of existing ones. Consequently, the static nature of the indexing mechanism proves inadequate for addressing real-time updates, thereby necessitating alternative approaches to address this deficiency. In the subsequent section~\ref{sec:related}, we will delve into the strategies devised to mitigate the constraints associated with static graph-based indices.

\par HNSW (Hierarchical Navigable Small World) is a widely used and well-established graph-based index within the industry (\cite{8594636}), lauded for its exceptional performance in both search precision and computational efficiency. However, according to the findings and experimental analysis in our preliminary study (see details in section \ref{sec:preli}), 
it has come to light that the current iteration of the HNSW index manifests two notable shortcomings following a substantial sequence of delete, insert, and query operations. The first and primary concern arises from the utilization of the markDelete algorithm and the replaced\_update insertion algorithm (hereafter referred to as replaced\_update for brevity), leading to a scenario where specific data points may become inaccessible, a phenomenon we identify as the `unreachable points phenomenon.'

\par This phenomenon can pose problems in practical applications. Imagine two scenarios in typical systems: If certain data points representing online stores in an index become inaccessible after a series of modifications, such as insertions and deletions, it would be highly frustrating for the owners of those online stores. This is especially critical for store owners who rely on recommendation systems; if their stores can never be recommended due to the unreachable points phenomenon, despite being the most relevant to client requests, it would significantly impact their business. Similarly, in a RAG system, if a user seeks knowledge on a specific topic, they would include relevant keywords in their query. However, due to the unreachable points phenomenon, if the data points most relevant to those keywords become unsearchable, the accuracy of the retrieved results will be compromised, subsequently affecting the output of the generation model. Therefore, the problem of some points becoming unreachable after insertions and deletions is a real-world concern. The second issue is that we observed the efficiency of native HNSW operations involving mixed deletions and insertions is lower compared to query operations. This inefficiency leads to query delays in scenarios involving mixed deletions, insertions, and queries, indirectly reducing overall query efficiency. 

To address the first issue, we proposed practical solutions to mitigate this problem. Additionally, the deletion and update strategies we designed can reduce the number of unreachable points and their growth rate after repeated deletions and insertions. We redesigned the deletion and update strategies for HNSW to address the second issue. Our new method enhances the efficiency of mixed operations compared to the original approach and mitigates the impact of the phenomenon described in the first issue.
\par Our contribution can be summarized in the following:
\begin{itemize}
    \item \textbf{Revealing the unreachable point phenomenon in the HNSW index:}
    \begin{itemize}
        \item Through analysis and experimentation, we have identified that HNSW can create unreachable points in the graph after deletion and insertion operations, leading to adverse effects, as previously discussed. We proposed suitable practical solutions and mitigated the impact of this phenomenon in our newly designed deletion and update algorithm.
    \end{itemize}
    \item \textbf{Improved Replaced\_Update Strategy:}
    \begin{itemize}
        \item Building on existing methods, we proposed an improved replaced\_update algorithm called MN-RU. This algorithm enhances the speed of deletion and insertion operations while also alleviating the unreachable points phenomenon to a significant extent.
    \end{itemize}
    \item \textbf{Comprehensive Performance Evaluations:}
    \begin{itemize}
        \item We implemented a comprehensive strategy and integrated it into a single platform for comparative evaluations. The results validate the superiority of our approach over the native HNSW strategy, offering valuable insights for future practical applications.
    \end{itemize}
\end{itemize}

\section{Related Work}\label{sec:related}

\subsection{Approximate Nearest Neighbor Search Algorithms}
Approximate nearest neighbor search (ANNS) has been a central focus of scholarly research for the past two decades, which led many studies to develop effective methodologies to address this complex problem. 

In space partitioning methods, such as tree structures, data space is initially recursively divided into multiple regions to facilitate the construction of tree or forest-based indices. These methods are commonly categorized into hierarchical structures and reference point-based structures. While effective for low-dimensional datasets, their performance significantly deteriorates with increasing data dimensionality. Beyond 10 dimensions, tree-based space partitioning methods exhibit reduced efficiency, often slower than brute-force linear-scan methods. Strategies like the pyramid technique~\cite{7113317} and iDistance~\cite{jagadish2005idistance} have been proposed to combat the 'curse of dimensionality.'

Initially developed for addressing ANNS in the Hamming space~\cite{gionis1999similarity}, hash-based methods were later extended to the Euclidean space~\cite{datar2004locality}. These methods involve projecting high-dimensional data points into lower-dimensional spaces using hash functions to devise efficient algorithms to identify nearest neighbors. Despite their efficacy, hash-based methods often require numerous hash tables for satisfactory search outcomes, leading to the emergence of variants like VHP~\cite{lu2020vhp}, LCCS-LSH~\cite{lei2020locality}, PMLSH~\cite{zheng2020pm}, and R2SLH~\cite{lu2020r2lsh}.

Product quantization (PQ) is a widely used quantization-based method for accelerating search processing by compressing input vectors into compact codes for memory-efficient dataset processing. It also provides effective techniques for estimating distances between raw vectors and compressed codes, improving distance estimation accuracy. Diverse quantization-based approaches~\cite{babenko2014additive,babenko2015tree} have been devised to mitigate quantization errors and to improve query precision, focusing on refining quantization techniques.

Graph-based methods~\cite{li2019approximate,lee2022augmentation,oyamada2023meta,groh2022ggnn} have emerged as potent solutions for high-dimensional ANNS, constructing efficient indexing structures. However, constructing exact KNN graphs becomes exponentially complex with increasing nodes. Approximated KNN graph construction has been explored as an alternative, offering similar efficacy as exact graphs in specific applications. Despite this, memory constraints in graph construction necessitate cluster algorithms for simplified subgraph creation and more efficient query processing. Integration of machine learning and deep learning into graph-based nearest neighbor search methods has advanced search capabilities.

The landscape of dynamic update methods has recently advanced a lot. Noteworthy advancements include Fresh-diskann~(\cite{DBLP:journals/corr/abs-2105-09613}), which introduces a \textit{StreamingMerge} protocol to handle node deletions from the \textit{DeleteList} and subsequent insertion of $N$ new points, as detailed in the referenced paper. Additionally, SPANN~\cite{NEURIPS2021_299dc35e} stands out as the pioneering on-disk vector index, leveraging balanced clustering to achieve minimal tail search latency and deliver exceptional performance. Furthermore, SPFresh~(\cite{10.1145/3600006.3613166}) offers heightened throughput and reduced latency for both search and update operations, ensuring the efficient retrieval of new highly reliable vectors.

\subsection{HNSW}

Previous studies have investigated various strategies aimed at maintaining the freshness of graph-based or hybrid indexes. Notably, HNSW \footnote{\url{https://github.com/nmslib/hnswlib.git}} serves as a graph-based index in memory and has developed a robust framework for executing delete and update operations. In the case of a delete operation concerning a label $x_d$, HNSW employs the \textit{markDelete} algorithm to flag and transfer it to a designated deleted set. On the other hand, for an insertion operation, the choice can be made to either directly insert a new point into the index if the index size remains within the predefined limit, or to replace an old deleted label $x_d$ with the new entry. The whole insertion procedure, referred to as \textbf{replaced\_update} in this study, can be represented as follows. 
\begin{enumerate}
    \item \textbf{Check for Deleted Points}: Initially, HNSW examines for any points flagged as deleted. If a deleted point is identified, it is designated for replacement.
    \item \textbf{Collect Neighbors}:  The algorithm gathers the one-hop and two-hop neighbors of the deleted point. For each one-hop neighbor \( v_j \), the one-hop and two-hop neighbors and the newly inserted point are considered potential candidates new neighbors for \( v_j \).
    
    \item \textbf{Prune Candidate Neighbors}: Employing a pruning strategy, an optimal neighbor set \( N(v_j) \) is chosen for each one-hop neighbor \( v_j \) from the pool of candidates.
    
    \item \textbf{Update Connections}: Directed edges are established from each one-hop neighbor \( v_j \)  to the points in \( N(v_j) \), ensuring graph connectivity.
    
    \item \textbf{Insert the New Point}: The new point is integrated into the index, with connections established based on the rectified neighbor sets.
   
\end{enumerate}

\par The above-described processes are implemented within the HNSW source code, particularly encompassing the functions \textit{addPoint}, \textit{updatePoint}, and \textit{repairConnectionsForUpdate}. The significance of the \textbf{replaced\_update} method lies in its ability to manage the index efficiently. Without it, inserting new data points would leave deleted entries in the index, wasting storage and causing unnecessary expansion. By using \textbf{replaced\_update}, deleted points' storage is reused for new insertions, preventing the index from growing excessively and maintaining efficient space utilization.

Our approach extends from the foundational framework of HNSW, integrating enhancements to enhance the efficiency of dynamic updates and mitigate the issue of unreachable points. By optimizing the deletions and insertions, we aim to reduce latency and promote enhanced graph connectivity.



\section{Issues in Update Efficiency and Unreachable Points}\label{sec:preli}


\captionsetup{labelformat=empty} 
\begin{figure}[ht]
\centering
\includegraphics[width=0.4\textwidth]{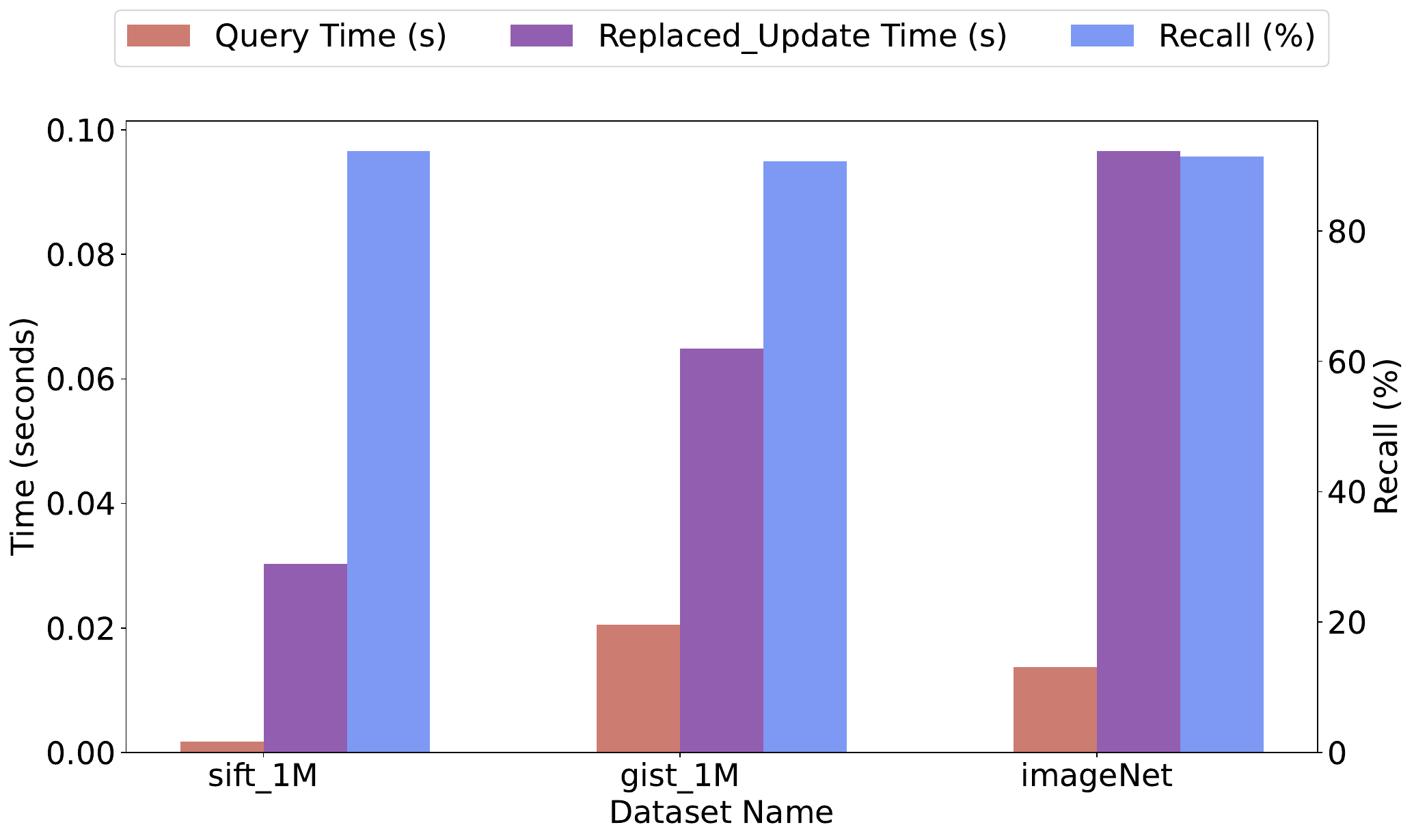}
\caption{Figure 1: Comparison of query efficiency and replaced\_update efficiency at a given recall level on three public datasets.}
\label{fig:query_vs_update}
\end{figure}

Although integrating the \textbf{replaced\_update} method in the HNSW algorithm to efficiently substitute deleted points with new entries, thus facilitating real-time point deletion and space reclamation, its operational efficiency remains subpar. Illustrated in Figure \ref{fig:query_vs_update}, when approaching a recall rate of approximately 90\%, the efficiency of \textbf{replaced\_update} lags notably behind the search efficiency of the query in three public datasets. Notably, in datasets like GIST and ImageNet, the insertion speed using the replaced\_update method is notably slower by a factor of 5 to 10. Through analyzing the process of \textbf{replaced\_update} method, it has come to our attention that this method may give rise to the `unreachable points phenomenon' after repeated deletions and insertions. That is, some specific points become inaccessible. Furthermore, this operational inefficiency compounds the issue of the unreachable points phenomenon. Subsequently, we will delve into the underlying reasons contributing to the occurrence of the unreachable points phenomenon and conduct an empirical experiment to validate our hypothesis.


\subsection{Causes of Unreachable Points}


\newtheorem{mydefinition}{Definition}
\begin{mydefinition}
An \textbf{unreachable point} is defined as a point that possesses outgoing edges but lacks incoming edges in \textbf{all} layers of the navigable small-world graph.
\end{mydefinition}

In Figure \ref{fig:unreachable_points_phenomenon_demon}, node $v$ has only one incoming edge from node $d$. After marking node $d$ for deletion and applying the HNSW replaced\_update strategy to remove $d$, node $v$ lacks any incoming edges. In other words, unless the node $v$ serves as the entry point in the HNSW structure, it will remain unvisited in subsequent search operations.

\captionsetup{labelformat=empty} 
\begin{figure}[h]
\centering
\includegraphics[width=0.5\textwidth]{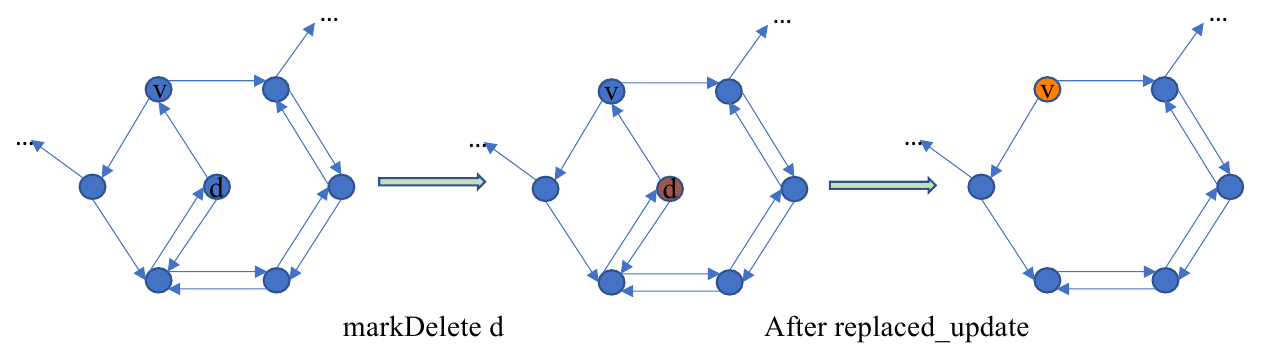}
\caption{Figure 2: Example Of Unreachable Points Phenomenon}
\label{fig:unreachable_points_phenomenon_demon}
\end{figure}

\par It is evident that a point with an in-degree of zero cannot be found during the search process. During the construction of the HNSW index, the likelihood of forming unreachable points is exceedingly small. However, if the replace\_update algorithm used by HNSW is applied to replace old points with newly inserted points, unreachable points emerge. 

\par In HNSW, the occurrence of some points having an in-degree of zero after the replaced\_update operation is attribute to the specific repair connection strategy employed by the algorithm. During a replaced\_update in HNSW, the algorithm first gathers the neighbors of the deleted point, referred to as \(T\). It then combines \(T\) with their own neighbors and the newly inserted point to form a new candidate neighbor set for each member of \(T\). Finally, new neighbors are selected from the candidate set for each member of \(T\). The edges from members of \(T\) to their original neighbors might be deleted during these operations. As a result, these original neighbors may end up with no in-edges after the replaced\_update operation. Consequently, these points will not be found during future search processes unless they are the entry points of the HNSW. 

\begin{figure}[h]
    \centering
    \begin{minipage}[t]{0.24\textwidth}
        \centering
        \includegraphics[width=\textwidth]{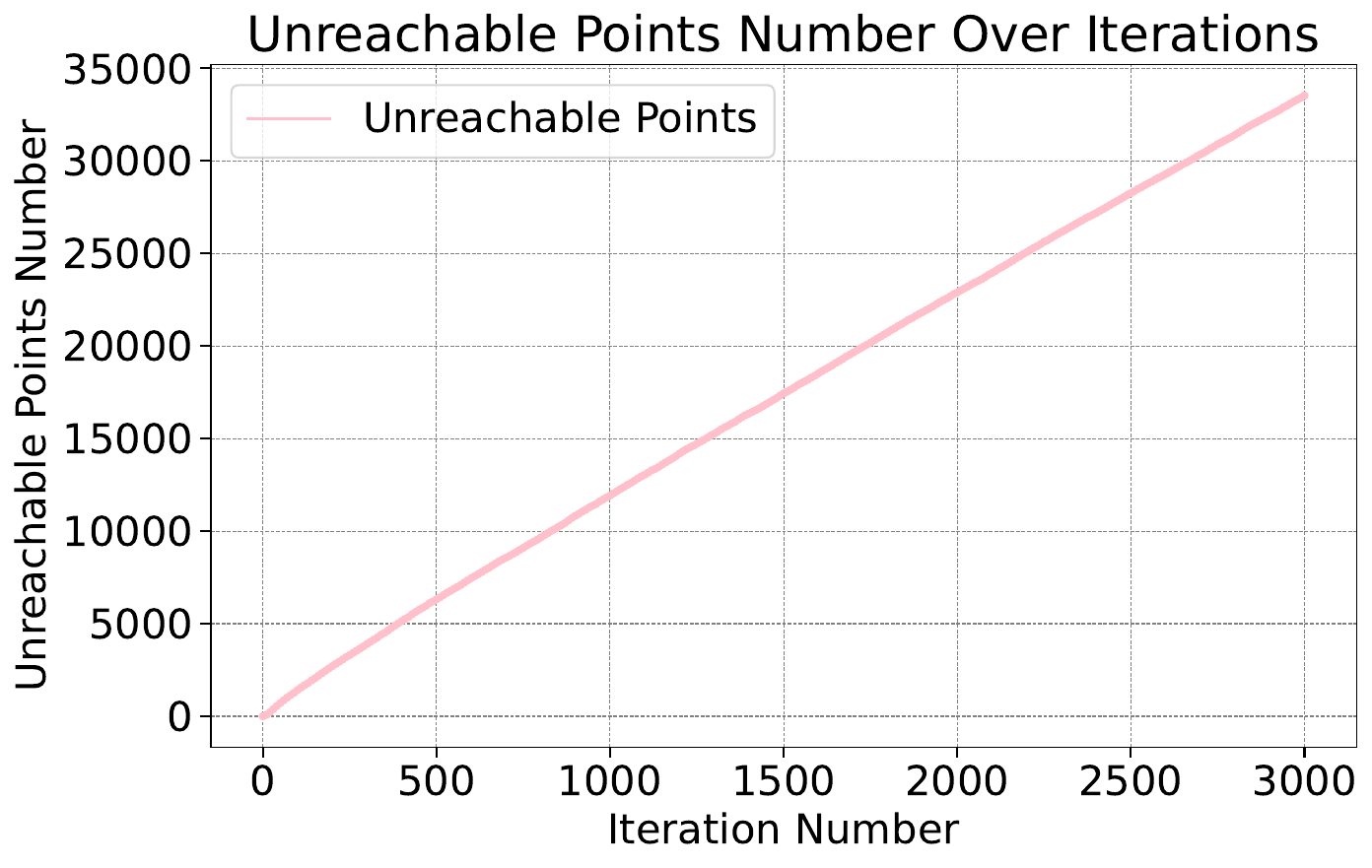}
        \caption*{(a) Unreachable points number over iterations}
    \end{minipage}
    \begin{minipage}[t]{0.24\textwidth}
        \centering
        \includegraphics[width=\textwidth]{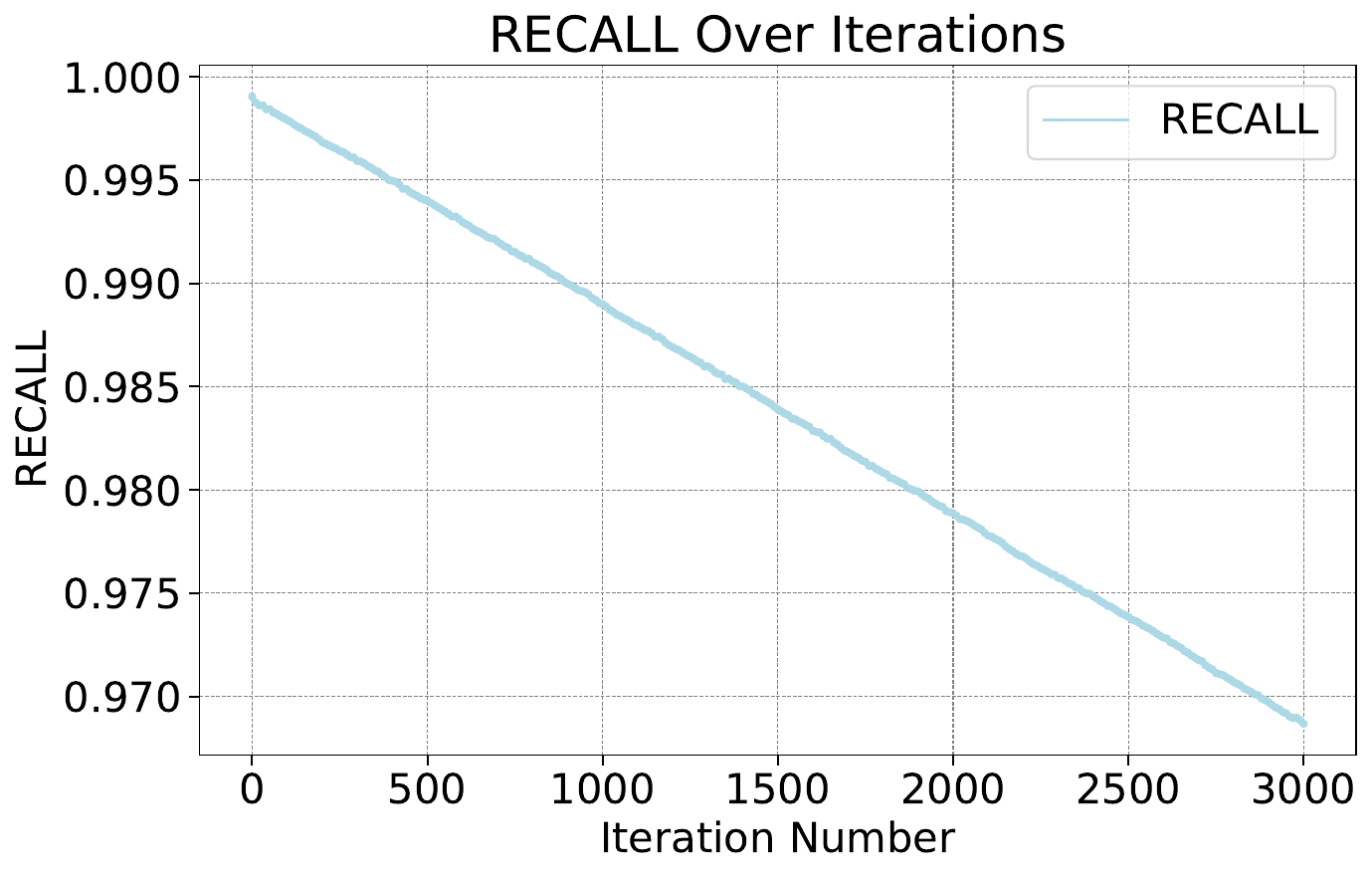}
        \caption*{(b) RECALL over iterations}
    \end{minipage}
    \caption{Figure 3: The demonstration of the unreachable points phenomenon on the Sift dataset. In each iteration, 5\% of the data points are randomly deleted from the Sift dataset, while ensuring that none of the deleted points were previously unreachable. These points are then reinserted back.}
    \label{fig:unreachable_points_phenomenon}
\end{figure}

\subsection{Demonstration on Specific Datasets}



\par Figure \ref{fig:unreachable_points_phenomenon} illustrates the presence of unreachable points in HNSW following replaced\_update operations. The experiment uses the Sift dataset, where 5\% of the data is randomly deleted in each iteration, ensuring that these points do not already belong to the set of unreachable points and are subsequently reinserted. The experiment consists of 3000 iterations to track the evolution of unreachable points after each cycle of deletion and insertion.

\par Figure \ref{fig:unreachable_points_phenomenon} (a) illustrates that the number of unreachable points in HNSW consistently increases with continuous deletion and reinsertion. After approximately 3000 iterations, the number of unreachable points reaches between 3\% and 4\% of the original dataset. This number rises with additional iterations, indicating that points already in the unreachable points set will not be searched in future search processes.

\par Figure \ref{fig:unreachable_points_phenomenon} (b) shows that recall decreases as the number of iterations increase. With the same configuration, recall declines by about 3\% due to the increasing number of unreachable points over iterations. This decline becomes more severe with more iterations, leading to a gradual reduction in the accuracy of the final search. Furthermore, this decrease in accuracy caused by the unreachable points phenomenon cannot be mitigated by adjusting HNSW searching parameters, such as increasing the ef\_ parameter.




\captionsetup{labelformat=empty} 
\begin{figure*}[htbp]
\centering
\includegraphics[width=0.55\textwidth]{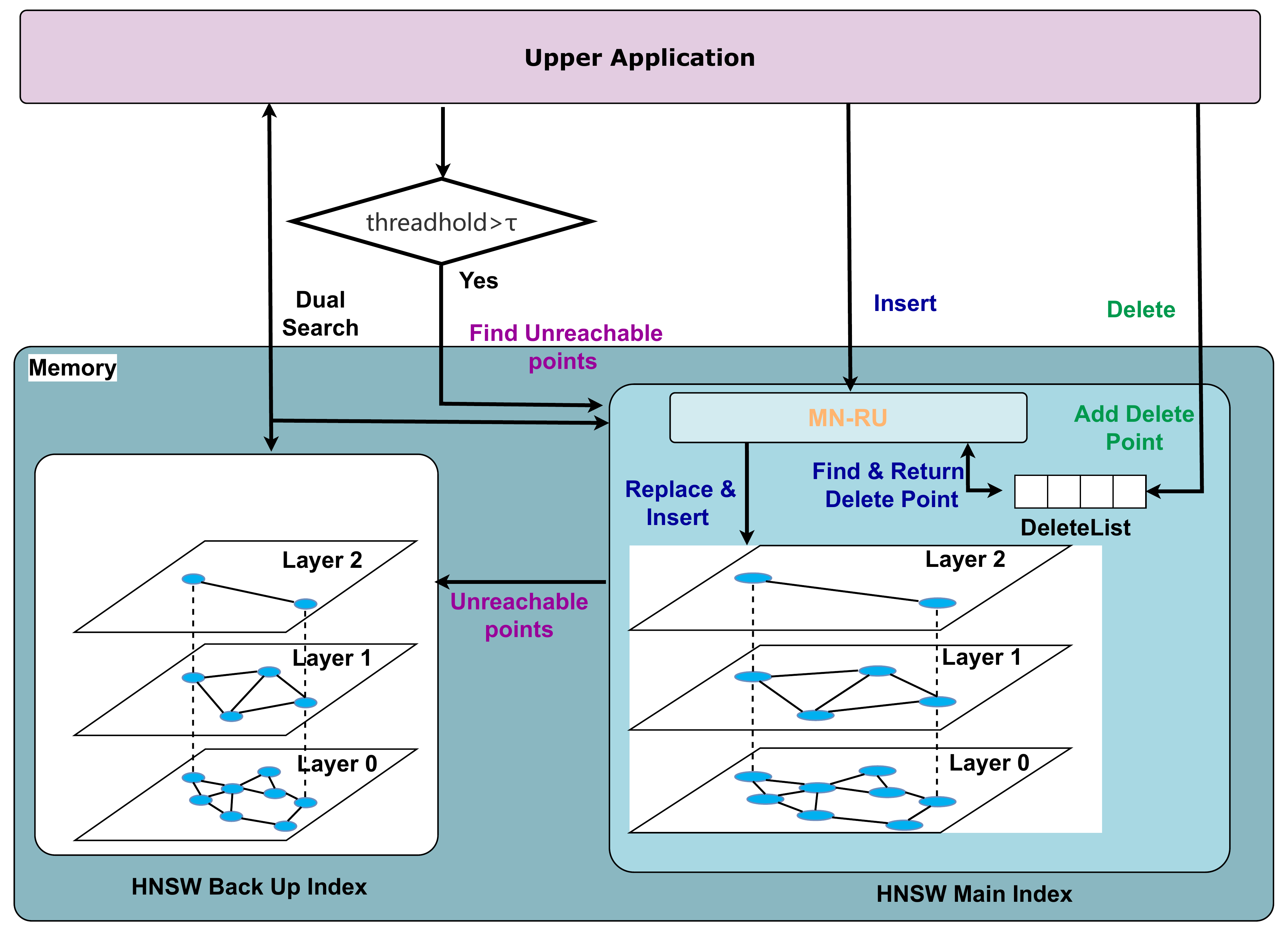}
\caption{Figure 4: The architecture of our work, both in upper-level application and MN-RU}
\label{fig:architecture}
\end{figure*}

\section{Methodology}

As shown in Figure \ref{fig:query_vs_update}, the HNSW replaced\_update algorithm exhibits poor update performance, leading to the phenomenon of unreachable points, which adversely affects search performance, as illustrated in Figure \ref{fig:unreachable_points_phenomenon}. To address these issues, we propose the back\_up\_index\_construction and dual\_search methods to mitigate the growth of unreachable points from the upper-level application. More importantly, we introduce the MN-RU algorithm, designed to improve update efficiency and significantly reduce the incidence of unreachable points within the index. The architecture of our approach is shown in Figure \ref{fig:architecture}, which illustrates the integration of the HNSW index in the upper-level application and the operation of the MN-RU within the HNSW index.

\subsection{Dual Indexes Design}

In an HNSW index, points becoming unreachable after replaced\_update operations can be attributed to the graph's inherent connectivity and the algorithm's maintenance processes. Although reconstruction and reconnection can effectively restore the connectivity of the HNSW graph, they also present several drawbacks, such as significant computational resources and time, service interruption, and system complexity.
In practical applications involving high-frequency update scenarios, it is imperative to weigh these disadvantages carefully. This article considers strategies to address the point-disconnected problem by maintaining graph connectivity to balance performance and availability.

Diverging from conventional methodologies reliant on a singular HNSW index, we introduce an additional index designed explicitly for managing unreachable points within the HNSW indexing system, referred to as the \textit{HNSW Backup Index}, as illustrated in Figure \ref{fig:architecture}. The inception of this algorithm is rooted in minimizing the computational burden and service disruptions associated with recurrent reindexing processes. The procedure for creating the \textit{Backup Index} involves several steps: First, the HNSW K-NN search function is applied to identify the point set \(F\) that is closest to the query point \(q\) from the current data set \(P\). We set \(K\) to \(|P|\) to retrieve all the points available in the index. Next, by removing \(F\) from the set \(P\), the remaining set of points \(U\), which includes the points not found, is identified and extracted. Subsequently, a new HNSW backup index is constructed for these unfound point sets \(U\). This process leverages the multi-level proximity search feature of HNSW to efficiently manage unreachable points, ensuring that the structure and query performance of the original index remains unaffected while reducing the computational overhead and service interruptions typically associated with reconstruction. 

Meanwhile, we introduce a threshold \(\tau\) to regulate replaced\_update operations. When the number of replaced\_update operations exceeds \(\tau\), it triggers the reconstruction of the \textit{HNSW backup index}. The upper application layer can adjust the value of \(\tau\) according to specific needs. In our implementation, the value of \(\tau\) was empirically configured to 40000.
 
Based on this novel index structure, it is imperative to propose new query and maintenance strategies to maintain the accuracy of the results, optimize query performance, and prevent any potential degradation in system efficiency. Next, we will describe these strategies comprehensively. 


\subsection{Dual Index Search}

In order to ensure the query efficiency and the accuracy of the results, we propose a query algorithm called \textit{dualSearch}, which queries a primary index (primary HNSW index) and a backup index (HNSW index dedicated to managing unreachable points) simultaneously to ensure that the query can cover all possible data points, even if some points no longer appear in the primary index after the replaced\_update operations.

Algorithm~\ref{alg:dual_search} outlines the search process for \textit{dualSearch}. Initially, it performs a K-NN search on the primary index (\(HNSW_{main}\)) using the query point \(q\) to retrieve \(k\) nearest neighbors in the set of results \(R_{main}\). Concurrently, a similar K-NN search is executed on the backup index (\(HNSW_{backup}\)), dedicated to managing unreachable points, yielding the results set \(R_{backup}\). Subsequently, the results of both searches are merged into a unified set \(R_{combined}\), which is then sorted according to the distance from point to query to prioritize closer points. Ultimately, the algorithm delivers the top \(k\) points from this ordered set, ensuring a precise and effective query result that takes into account both the accessible and the unreachable points.

\vspace{-0.75em}
\begin{algorithm}\scriptsize
\caption{dualSearch(q, k, j)}
\begin{algorithmic}[H]
\STATE \textbf{Input:} Query point: $q$, number of neighbors: $k$, size of dynamic candidate list: $ef$
\STATE \textbf{Output:} Top $k$ points from combined search results
\STATE \textbf{Variables:} $R_{main}$: Result set from main index, $R_{backup}$: Result set from backup index, $R_{combined}$: Combined result set, $R_{sorted}$: Sorted result set
\STATE $R_{main} \leftarrow$ K-NN-SEARCH$(HNSW_{main}, q, k, ef)$ // K-NN-SEARCH is Algorithm 5 from HNSW
\STATE $R_{backup} \leftarrow$ K-NN-SEARCH$(HNSW_{backup}, q, k, ef)$ // K-NN-SEARCH is Algorithm 5 from HNSW
\STATE $R_{combined} \leftarrow R_{main} \cup R_{backup}$ // Combine results
\STATE $R_{sorted} \leftarrow$ sort $R_{combined}$ by distance to $q$
\RETURN Top $k$ points from $R_{sorted}$
\end{algorithmic}
\label{alg:dual_search}
\end{algorithm}
\vspace{-0.75em}

This approach improves search accuracy and maintains robust query performance, even with unreachable points. By using a dedicated backup index, the system avoids the pitfalls of traditional single-index structures.

\subsection{Index Maintenance}
\label{Index Maintenance}

When confronted with extensive datasets that undergo frequent updates through insertions and deletions, ensuring the precision of query outcomes requires an efficient strategy to maintain the index that encompasses update, delete, and other operations. This challenging task is pivotal for maintaining the integrity and reliability of the query results.  

As mentioned earlier, the original HNSW replaced\_update algorithm needs to consider the neighbors of the deleted point(namely one-hop neighbors) and the neighbors of these neighbors (namely two-hop neighbors) as potential new neighbors to one-hop neighbors during the update process. For each one-hop neighbors, the algorithm has to reselect its neighbors from a candidate set of size $M^2$, including one-hop neighbors and the two-hop neighbors. Here, $M$ is a parameter in HNSW that defines the maximum out-degree of a data point within a specific layer. The time complexity of this operation is $O(M^3)$ per layer, which is significantly time-consuming.

\begin{figure}
    \centering
    \includegraphics[width=0.5\textwidth]{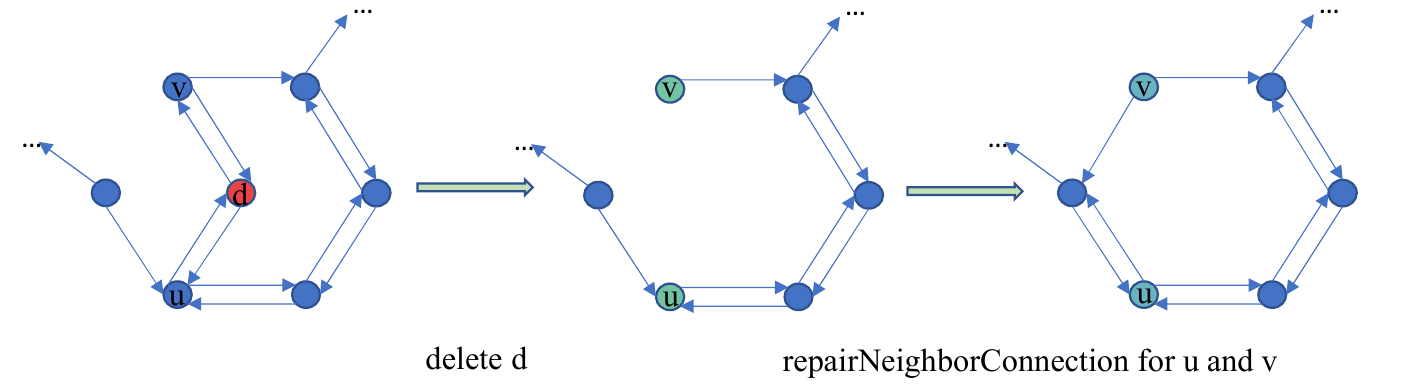}
    \caption{Figure 5: Example of repairing neighbors of points u and v after deleting point d.}
    \vspace{-1em}
    \label{fig:repair}
\end{figure}

For selecting new neighbors for one-hop neighbors, considering all two-hop neighbors as candidates is inefficient. The reason is as follows: Before update, based on HNSW's edge selection strategy, if a one-hop neighbor does not have an edge to a particular two-hop neighbor, it means HNSW deemed this edge suboptimal. Since the size of two-hop neighbors is $M^2$ and each one-hop neighbor originally has $M$ neighbors, many two-hop neighbors are not part of its original neighbor set. Therefore, establishing edges from one-hop neighbor to such two-hop neighbors was deemed suboptimal by HNSW before update and is likely to remain so after update, as the index structure does not undergo significant changes.

To address this complexity, we propose a novel algorithm(Algorithm \ref{alg:repairNeighborConnection}) which focuses on restoring the connectivity of one-hop neighbors of the deleted point. Contrary to original method, Algorithm \ref{alg:repairNeighborConnection} exclusively updates neighbors of one-hop neighbors that include the deleted point in their original neighbor list. For every such point, the candidate new neighbors comprise its original neighbors along with the neighbors of the deleted point. Due to the previously described reasons, two-hop neighbors are not necessarily included in the new neighbor candidate set, ensuring a more efficient and focused neighbor update process. This operation exhibits a time complexity of $O(M^2)$ per layer since for each one-hop neighbors, the algorithm has to reselect its neighbors from a candidate set of size $M$. Therefore, our algorithm demonstrates superior efficiency compared to the original approach. Figure \ref{fig:repair} serves as an illustrative example of Algorithm \ref{alg:repairNeighborConnection}. The complete process of Index Maintenance is detailed below:

\underline{\textbf{Deletion}}. For deletions, as illustrated in Figure \ref{fig:architecture}, when a point in the index is deleted, our method appends it to a list, \textit{deletedList}, similar to the process in the original HNSW. The purpose of utilizing the \textit{deletedList} is to manage deleted points without immediate removal efficiently, thus preserving the graph's structural integrity during updates. 

\vspace{-0.75em}
\begin{algorithm}[H]\scriptsize
\caption{MNRepairNeighborConnection($data$, $label$, $\alpha$, $hnsw$)}
\begin{algorithmic}[1]
\STATE \textbf{Input:} New point data: $data$, label of the new point: $label$, multilayer graph: $hnsw$, parameter for pruning function: $\alpha$
\STATE \textbf{Variables:} 
\STATE \hspace{1em} $deletedPoint$: Point marked as deleted
\STATE \hspace{1em} $L_{max}$: Maximum layer of $deletedPoint$ in $hnsw$
\STATE \hspace{1em} $N_1$: Neighbor set 1
\STATE \hspace{1em} $N_2$: Neighbor set 2
\STATE \hspace{1em} $C$: Combined candidate set
\STATE $deletedPoint \leftarrow$ getDeletedPoint($hnsw$) // Get a point marked as deleted
\IF{$deletedPoint$ is NULL}
    \STATE insert($data$, $label$, $hnsw$) // Perform normal insertion
    \RETURN
\ENDIF

\STATE $L_{max} \leftarrow$ getMaxLayer($deletedPoint$, $hnsw$) // Get the maximum layer of the deleted point in $hnsw$
\FOR{$layer \leftarrow 0$ to $L_{max}$}
    \STATE $N_1 \leftarrow$ getNeighborhood($deletedPoint$, $layer$) // Get the neighbors of deletedPoint
    \STATE $N_2 \leftarrow \emptyset$ // Initialize empty set for $N_2$
    \FOR{each point $v \in N_1$}
        \IF{edge($v$, $deletedPoint$) exists in $layer$}
            \STATE $N_2 \leftarrow N_2 \cup \{v\}$ // Add $v$ to $N_2$
        \ENDIF
    \ENDFOR
    \FOR{each point $u \in N_2$}
        \STATE $C \leftarrow$ getNeighborhood($u$, $layer$) // Get neighbors of $u$
        \STATE $C \leftarrow C \cup N_1 \cup label$  // Combine candidate sets, set new insert point to candidates
        \STATE $C \leftarrow$ pruneCandidates($C$, $layer$, $\alpha$) // Prune the candidates using $\alpha-RNG$
        \STATE setNeighbors($u$, $C$, $layer$) // Set $u$'s neighbors to $C$
    \ENDFOR
\ENDFOR
\STATE update($deletedPoint$, $data$, $label$) // Update the data of the deleted point
\end{algorithmic}
\label{alg:repairNeighborConnection}
\end{algorithm}
\vspace{-0.75em}

\underline{\textbf{Insertion}}. Algorithms \ref{alg:repairNeighborConnection} and \ref{alg:update}, collectively called Mutual Neighbor Replaced Update (\textbf{MN-RU}), update the HNSW index when a new point replaces a deleted one. First, Algorithm \ref{alg:repairNeighborConnection} repairs the graph's connectivity, then Algorithm \ref{alg:update} inserts the new point to the index. Unlike the conventional approach of adding new points, the Algorithm \ref{alg:update} enables the new point to inherit the layer level of the deleted point. Subsequent to this inheritance, the new point undergoes insertion using the standard HNSW insert process. Algorithm \ref{alg:update} starts by identifying the top layer of the HNSW index and the maximum layer of the deleted point, establishing an initial entry point for the search. From the top layer to the layer above the deleted point's maximum layer, it iteratively searches for the nearest point as the entry point to the new data point, updating the entry point each time. Then, from the maximum layer of the deleted point to the lowest layer, it searches for neighbors, selects neighbors for the new point, and establishes connections. MN-RU ensures the seamless integration of the new element into the HNSW structure.

\section{Experiments}

In this section, we evaluate our algorithm and highlight key experimental observations. The implementations of our algorithm and other baseline methods were executed in C++ on a system equipped with an Intel Xeon CPU E5-2678 v3 @ 2.50GHz and 110GB memory, operating on Ubuntu 22.04. All computational tasks, like data insertion, deletion, and retrieval, utilized 40 threads for enhanced efficiency in this section.

\vspace{-0.75em}
\begin{algorithm}[H]\scriptsize
\caption{update(deletedPoint, data, label)}
\begin{algorithmic}[1]
\STATE \textbf{Input:} The data which is marked deleted deletedPoint, Data of the new element $data$, Label of the new element $label$
\STATE \textbf{Variables:}
\STATE \hspace{1em} $L_{max}$: Maximum layer of the deleted point
\STATE \hspace{1em} $L$: Maximum layer of the current HNSW
\STATE \hspace{1em} $ep$: Entry point for search
\STATE \hspace{1em} $W$: List of currently found nearest elements
\STATE \hspace{1em} $neighbors$: Neighbor points of current point
\STATE $L_{max} \leftarrow$ getMaxLayer($deletedPoint$) // Get maximum layer of the deleted point
\STATE $L \leftarrow$ getMaxLayer($hnsw$) // Get maximum layer of HNSW
\STATE $ep \leftarrow$ getEnterPoint($L$) // Get entry point for the highest layer of HNSW
\FOR{$lc \leftarrow L$ to $L_{max} + 1$}
    \STATE $W \leftarrow$ SEARCH-LAYER($data$, $ep$, 1, $lc$) // Search layer to find nearest elements
    \STATE $ep \leftarrow$ getNearestElement($W$, $data$) // Update entry point for next layer
\ENDFOR
\FOR{$layer \leftarrow L_{max}$ to $0$}
    \STATE $W \leftarrow$ SEARCH-LAYER($data$, $ep$, $efConstruction$, $layer$) // Search layer to find nearest elements
    \STATE $neighbors \leftarrow$ SELECT-NEIGHBORS($data$, $W$, $M$, $layer$) // Select neighbors for the current layer
    \STATE addBidirectionalConnections($neighbors$, $data$, $layer$) // Add bidirectional connections using the same strategy as HNSW
    \STATE $ep \leftarrow$ $W$ // Update entry point for next layer
\ENDFOR
\end{algorithmic}
\label{alg:update}
\end{algorithm}
\vspace{-0.75em}

\subsection{Datasets}

We used four datasets with different sizes and dimensions,
as shown in Table \ref{tab:data_statistics}. These datasets are widely
used for the evaluation of various ANNS methods, and we accessed these datasets through a public repository\cite{li2018general}~\footnote{\url{https://www.cse.cuhk.edu.hk/systems/hash/gqr/datasets.html}}. The Sift2M dataset is a subset comprising the first two million vectors extracted from the Sift1B dataset \cite{amsaleg2010datasets}, which encompasses 1 billion SIFT descriptors with a dimensionality of 128.




\begin{table}[H]\scriptsize
\centering
\caption{Table 1: Data statistics}
\label{tab:data_statistics}
\begin{tabular}{ccccc}
\toprule
\textbf{Dataset} & \textbf{Sift} & \textbf{Gist} & \textbf{ImageNet} & \textbf{Sift2M} \\
\midrule
Base Size & 1,000,000 & 1,000,000 & 2,340,373 & 2,000,000  \\
Dim & 128 & 960 & 150 & 128 \\
\bottomrule
\end{tabular}
\end{table}


\subsection{Baselines}



Given that most memory-based ANNS methods \cite{10.14778/3303753.3303754,10.14778/3489496.3489506,9383170} do not support real-time update operations, our primary comparison centers on contrasting our methods with HNSW's replaced\_update algorithm. Our methods, based mainly on Algorithm \ref{alg:repairNeighborConnection}, focus on reselecting neighbors only for points mutually connected with the deleted points, except for MN-THN-RU, which will be discussed later. These methods are collectively termed Mutual Neighbor Replaced Update (MN-RU) with distinctive suffixes. Hereafter, we introduce both the baseline methods and our strategy:




\begin{itemize}
        
    \item \textbf{HNSW replaced\_update algorithm(HNSW-RU)}: This method, derived from the original HNSW algorithm \cite{8594636}, functions by replacing deleted nodes within the index to uphold the structure and efficacy of the HNSW graph. The implementation of the replaced\_update algorithm can be accessed at \footnote{\url{https://github.com/nmslib/hnswlib.git}}. The HNSW index is constructed on four datasets. Specifically, for the SIFT and SIFT\_2M datasets \cite{amsaleg2010datasets}, the parameters are set at \( M = 16 \) and \( ef\_construction = 200 \). For the Gist dataset \cite{amsaleg2010datasets}, the parameters are \( M = 32 \) and \( ef\_construction = 600 \). In the case of the ImageNet dataset, the parameters are \( M = 64 \) and \( ef\_construction = 800 \). The remaining methods maintain consistent parameter settings.

    \item \textbf{Mutual Neighbor replaced\_update $\alpha$(MN-RU $\alpha$)}: This approach involves selecting the point set \( P \) with mutual connections to the deleted points for neighbor reselection. The candidate neighbor set \( C \) for \( P \) encompasses the neighbors of the deleted points and their neighbors. To enhance efficiency, the method incorporates the $\alpha$-RNG pruning strategy \cite{DBLP:journals/corr/abs-2105-09613}, with $\alpha = 1$.
    
    \item \textbf{Mutual Neighbor replaced\_update $\beta$(MN-RU $\beta$)}: Similar to the previous method, MN-RU $\beta$ selects the point set \( P \) with mutual connections to the deleted points for neighbor reselection. For each point \( v \) in \( P \), it includes the neighbors of the deleted points and \( v \)'s original neighbors as the candidate new neighbor set \( C \) for \(v\). The $\alpha$-RNG pruning strategy \cite{DBLP:journals/corr/abs-2105-09613} is applied with \( \alpha = 1\).
    
    \item \textbf{Mutual Neighbor replaced\_update $\gamma$ (MN-RU $\gamma$)}: MN-RU $\gamma$ is akin to MN-RU $\beta$ but with the $\alpha$ parameter of the $\alpha$-RNG pruning strategy set to 1.1 ($\alpha = 1.1$). This method directly implements Algorithm \ref{alg:repairNeighborConnection}.
    
    \item \textbf{Mutual Neighbor And Two Hop Neighbor replaced\_update (MN-THN-RU)}: A variant of MN-RU $\gamma$, MN-THN-RU reselect the neighbors for points that mutually connected with deleted points and neighbors of neighbors of deleted points that have a connection to the deleted point.

\end{itemize}


To control query recall in our experiment, we use $ef\_$ to represent the priority queue size in the HNSW search process, balancing query accuracy and efficiency. A larger $ef\_$ value results in higher recall but increased running time. Additionally, $K$ denotes the number of returned points.

\subsection{Experimental Scenarios And Metrics}


We compared our methods against the HNSW replaced\_update algorithm in the following three scenarios.

\begin{itemize}
    \item \textbf{Full\_Coverage}: In this scenario, encompassing the Gist, Sift, and ImageNet datasets, we execute 100 iterations where each dataset is segmented into 100 parts. Every iteration involves the deletion and reinsertion of a portion, enabling the assessment of the impact of complete coverage on the index structure and performance.
    
    \item \textbf{Random}: For the Gist, Sift, and ImageNet datasets, we conduct 200 iterations. Within each iteration, 10,000 labels are randomly generated for deletion and reinsertion, facilitating the evaluation of method performance and robustness in the face of random data manipulations.
    
    \item \textbf{New\_Data}: Focusing on the Sift2M dataset, we initialize the index with the first million data points and conduct 10 iterations. Each iteration deletes 100,000 points from the first million and inserts 100,000 points from the second million. This process aims to assess the index's performance and adaptability with continuous data introduction. The final index consists of the second million data points. 
\end{itemize}

\begin{figure*}[ht]
    \centering
    \begin{minipage}[t]{0.32\textwidth}
        \centering
        \includegraphics[width=\textwidth]{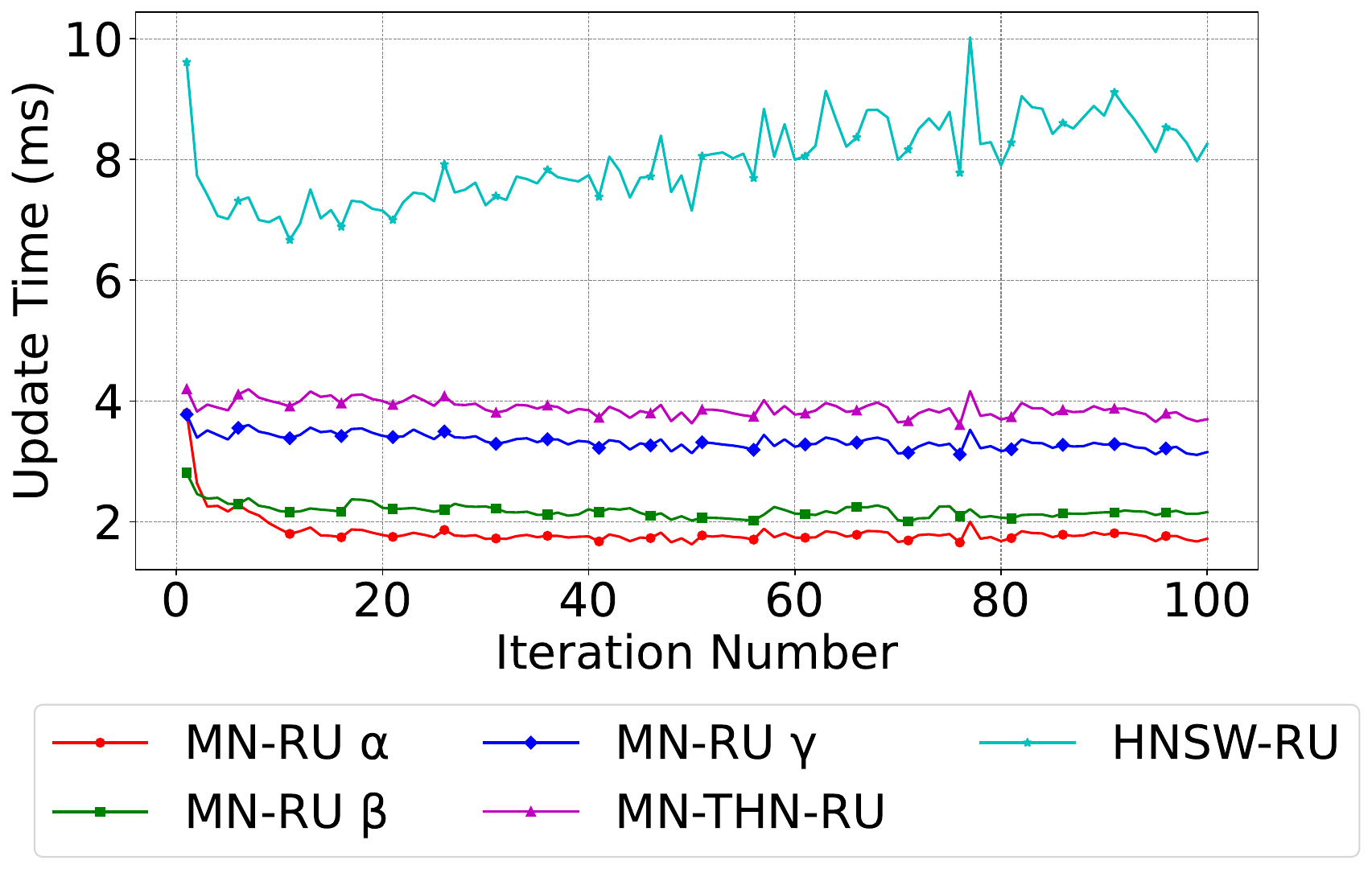}
        \caption*{(a) Gist}
    \end{minipage}
    \begin{minipage}[t]{0.32\textwidth}
        \centering
        \includegraphics[width=\textwidth]{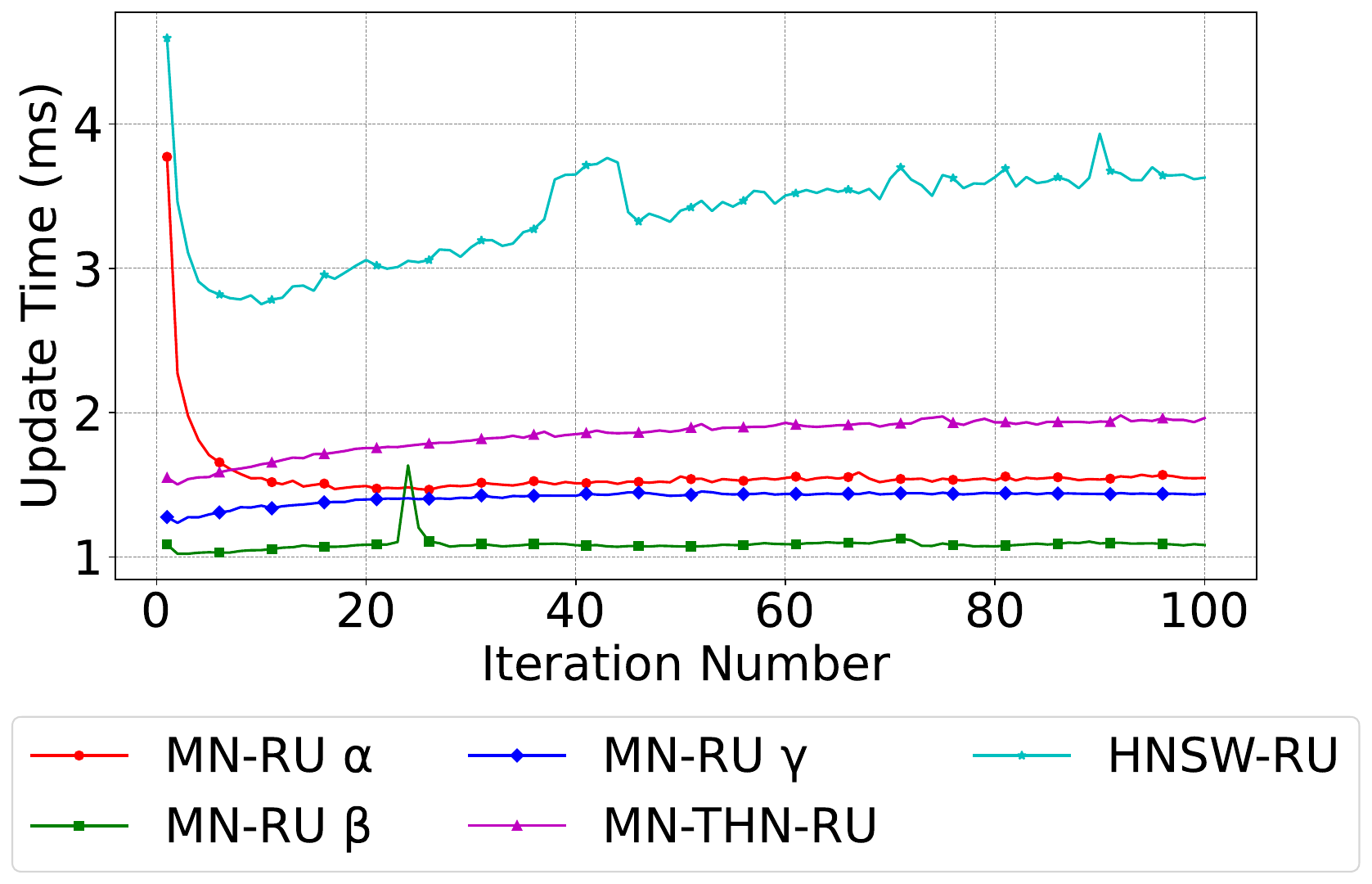}
        \caption*{(b) ImageNet}
    \end{minipage}
    \begin{minipage}[t]{0.32\textwidth}
        \centering
        \includegraphics[width=\textwidth]{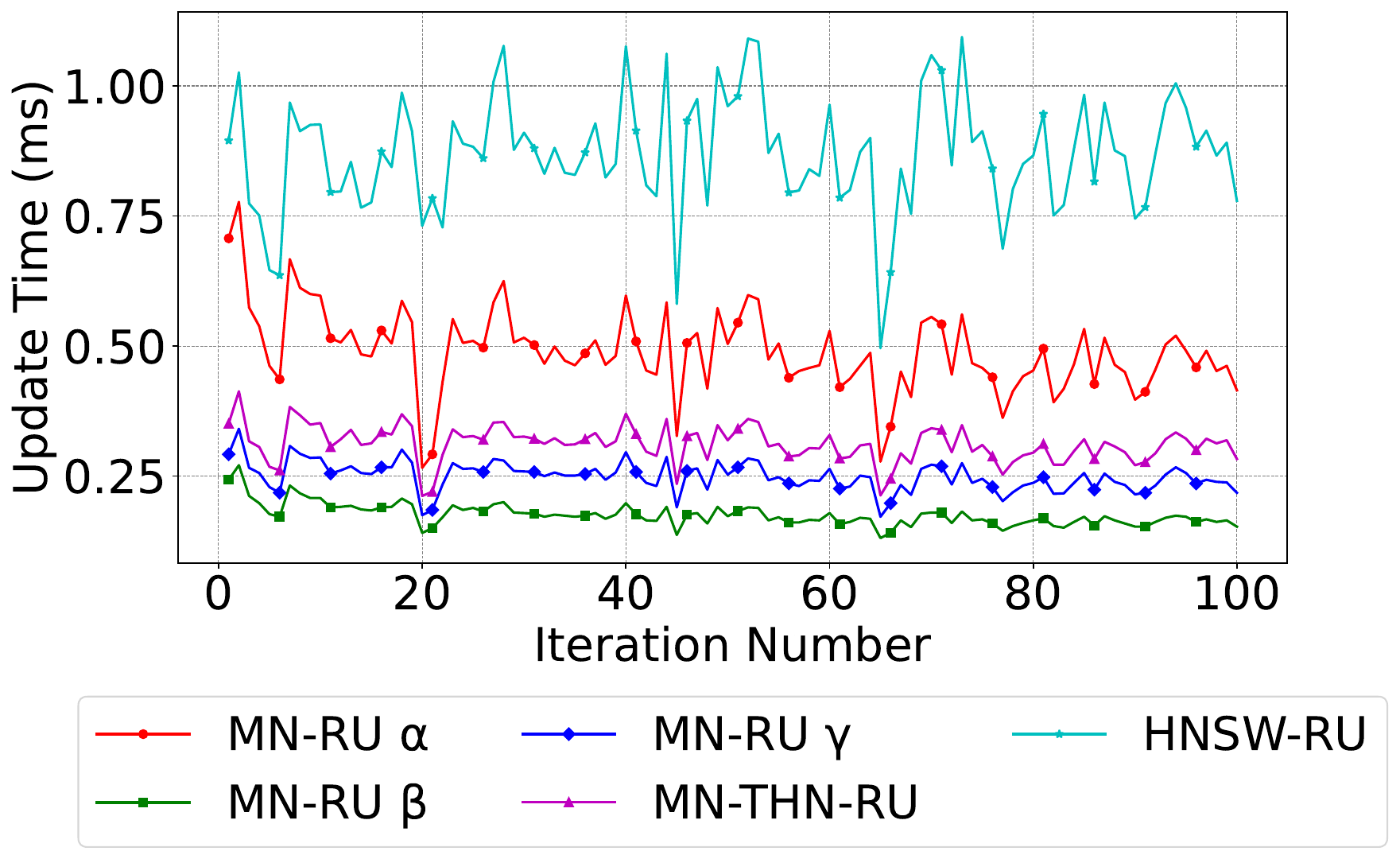}
        \caption*{(c) Sift}
    \end{minipage}
    \caption{Figure 6: Update time of different methods across various datasets in full\_coverage scenarios.}
    \label{fig:full_coverage_update_time}
    \vspace{-1em}
\end{figure*}

\begin{figure*}[ht]
    \centering
    \begin{minipage}[t]{0.32\textwidth}
        \centering
        \includegraphics[width=\textwidth]{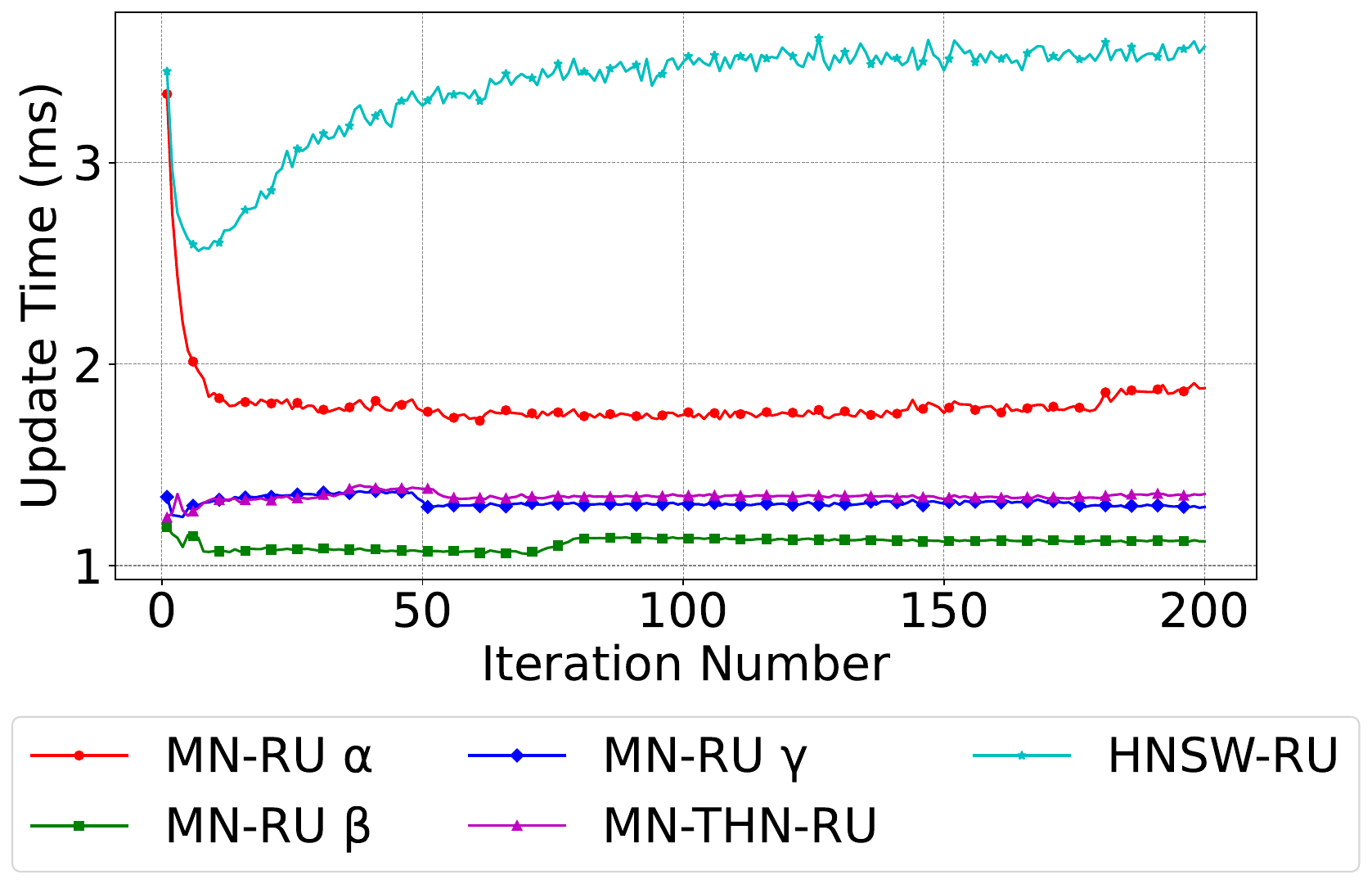}
        \caption*{(a) Gist}
    \end{minipage}
    \begin{minipage}[t]{0.32\textwidth}
        \centering
        \includegraphics[width=\textwidth]{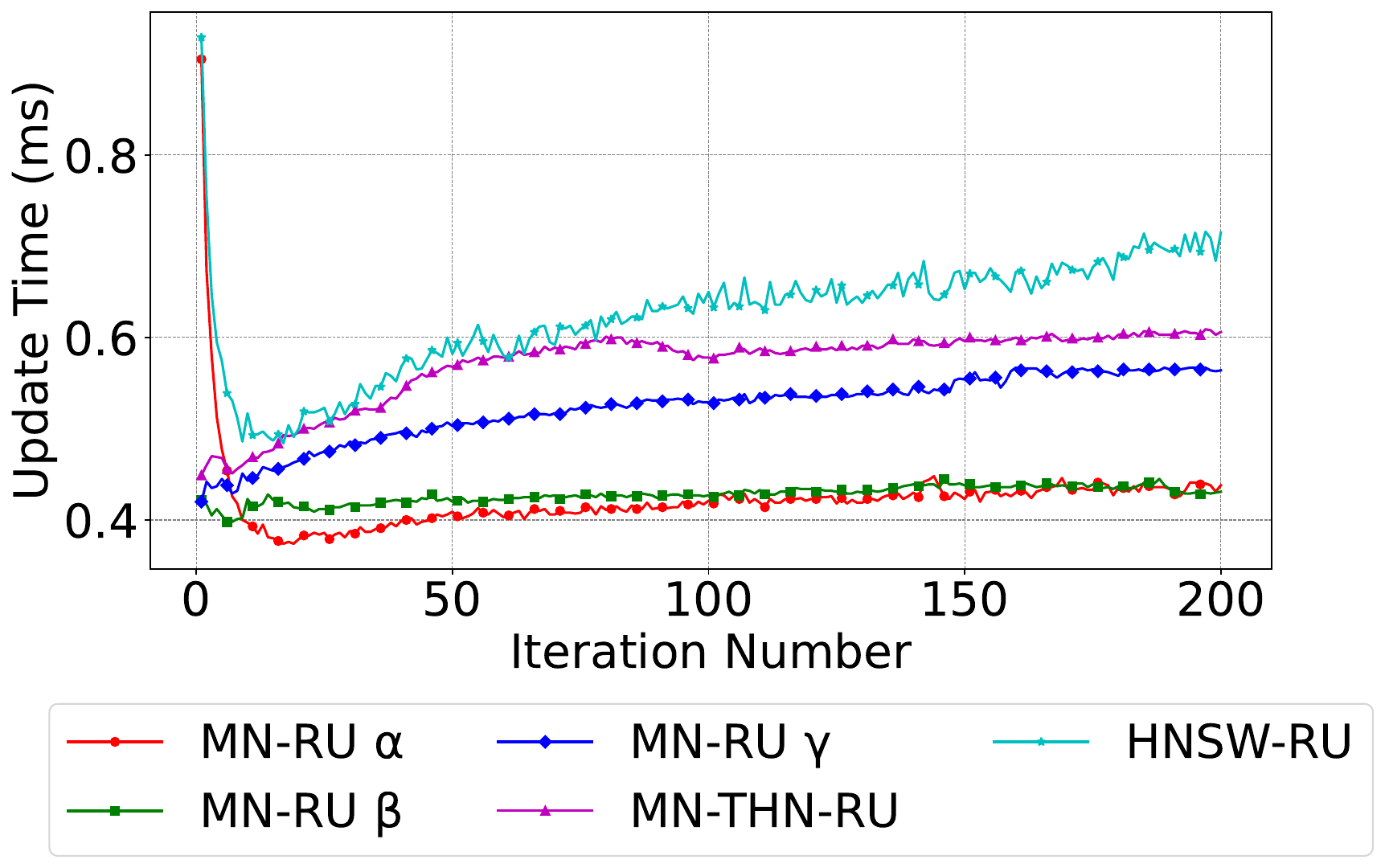}
        \caption*{(b) ImageNet}
    \end{minipage}
    \begin{minipage}[t]{0.32\textwidth}
        \centering
        \includegraphics[width=\textwidth]{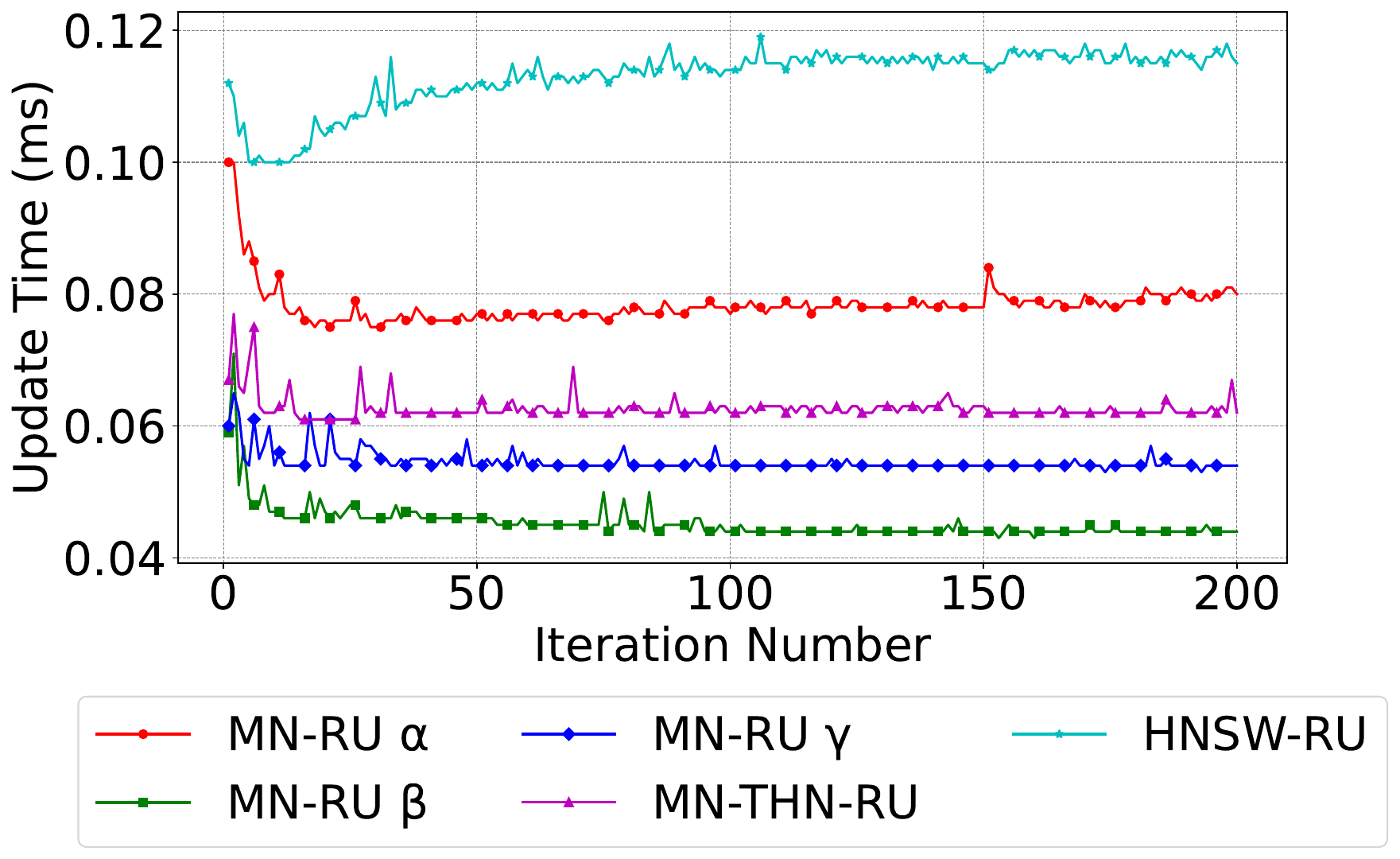}
        \caption*{(c) Sift}
    \end{minipage}
    \caption{Figure 7: Update time of different methods across various datasets in random scenarios.}
    \label{fig:random_update_time}
    \vspace{-2em}
\end{figure*}

This comparison relies on two key metrics: update time efficiency and growth of unreachable points. A method with superior efficiency exhibits reduced update times, while a method demonstrating excellence showcases minimal growth in unreachable points per iteration. We evaluated our methods against the HNSW replaced\_update algorithm in three scenarios, focusing on update time and the number of unreachable points.

\subsection{Update Time Efficiency}


This section compares our methods with native HNSW-RU methods regarding update time efficiency. The experiments were conducted in three scenarios described in the previous section: full\_coverage, random, and new\_data. The results are shown in Figures \ref{fig:full_coverage_update_time}, \ref{fig:random_update_time}, \ref{fig:new_data_update_time_unreachable_points} (a).


In the figures, the lower position of a curve corresponds to higher efficiency in update operations. It is evident that the MN-RU $\alpha$, MN-RU $\beta$, MN-RU $\gamma$, and MN-THN-RU methods consistently achieve lower update times compared to the HNSW-RU method and are 2-4 times faster in all scenarios. This indicates that our methods are more efficient in terms of update performance. The superior performance of our methods compared to HNSW-RU is due to their lower time complexity, as detailed in Section \ref{Index Maintenance}.



\subsection{Unreachable Points Growth}



In this section, we compare our methods with native HNSW-RU methods in terms of the growth of unreachable points. The experiments were conducted in three scenarios described in the previous section: full\_coverage, random, and new\_data. The results are shown in Figures \ref{fig:full_coverage_unreachable_points_growth}, \ref{fig:random_unreachable_points_growth}, and \ref{fig:new_data_update_time_unreachable_points} (b).


In the Figures, the method with fewer unreachable points after the same number of iterations is considered superior. We observe that over a prolonged period of update operations, the number of unreachable points increases for each method. For HNSW-RU, after 200 iterations described in previous sections, the number of unreachable points grows to approximately 3\% to 4\% in the Gist dataset and 2\% to 3\% in the ImageNet dataset. The MN-RU $\gamma$ and MN-THN-RU methods maintain fewer unreachable points than other methods, leading us to conclude that they are more effective.

\begin{figure*}[ht]
    \centering
    \begin{minipage}[t]{0.32\textwidth}
        \centering
        \includegraphics[width=\textwidth]{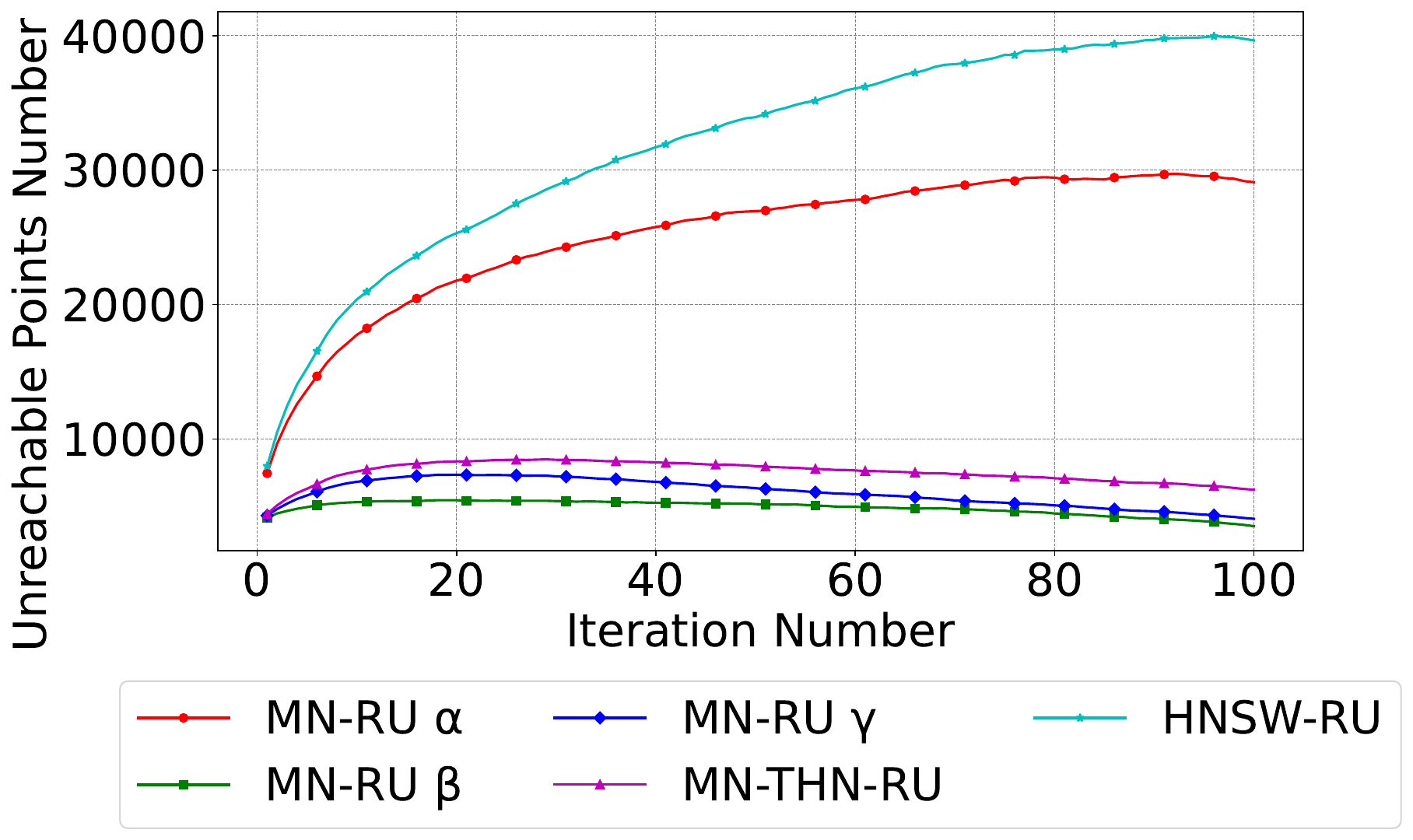}
        \caption*{(a) Gist}
    \end{minipage}
    \begin{minipage}[t]{0.32\textwidth}
        \centering
        \includegraphics[width=\textwidth]{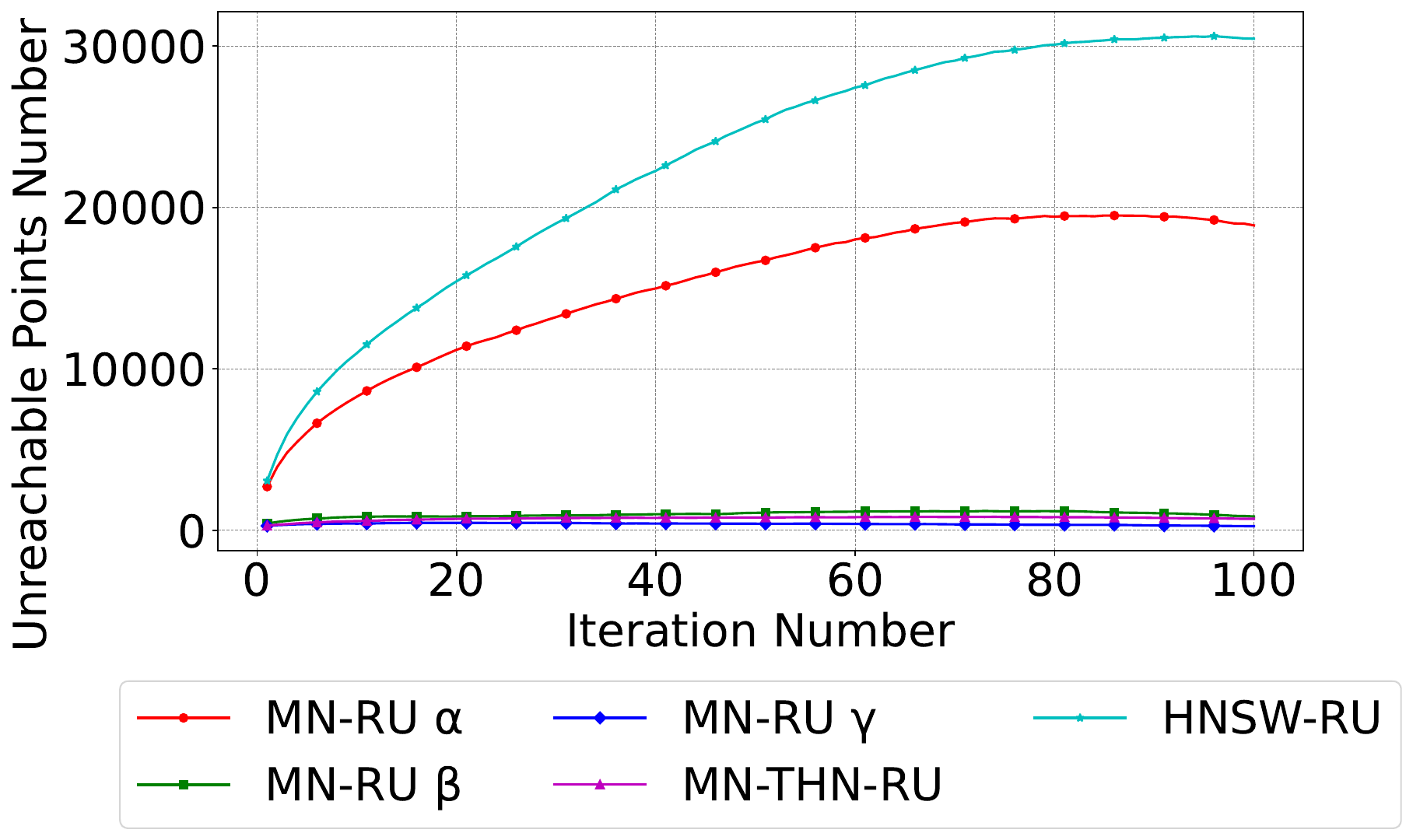}
        \caption*{(b) ImageNet}
    \end{minipage}
    \begin{minipage}[t]{0.32\textwidth}
        \centering
        \includegraphics[width=\textwidth]{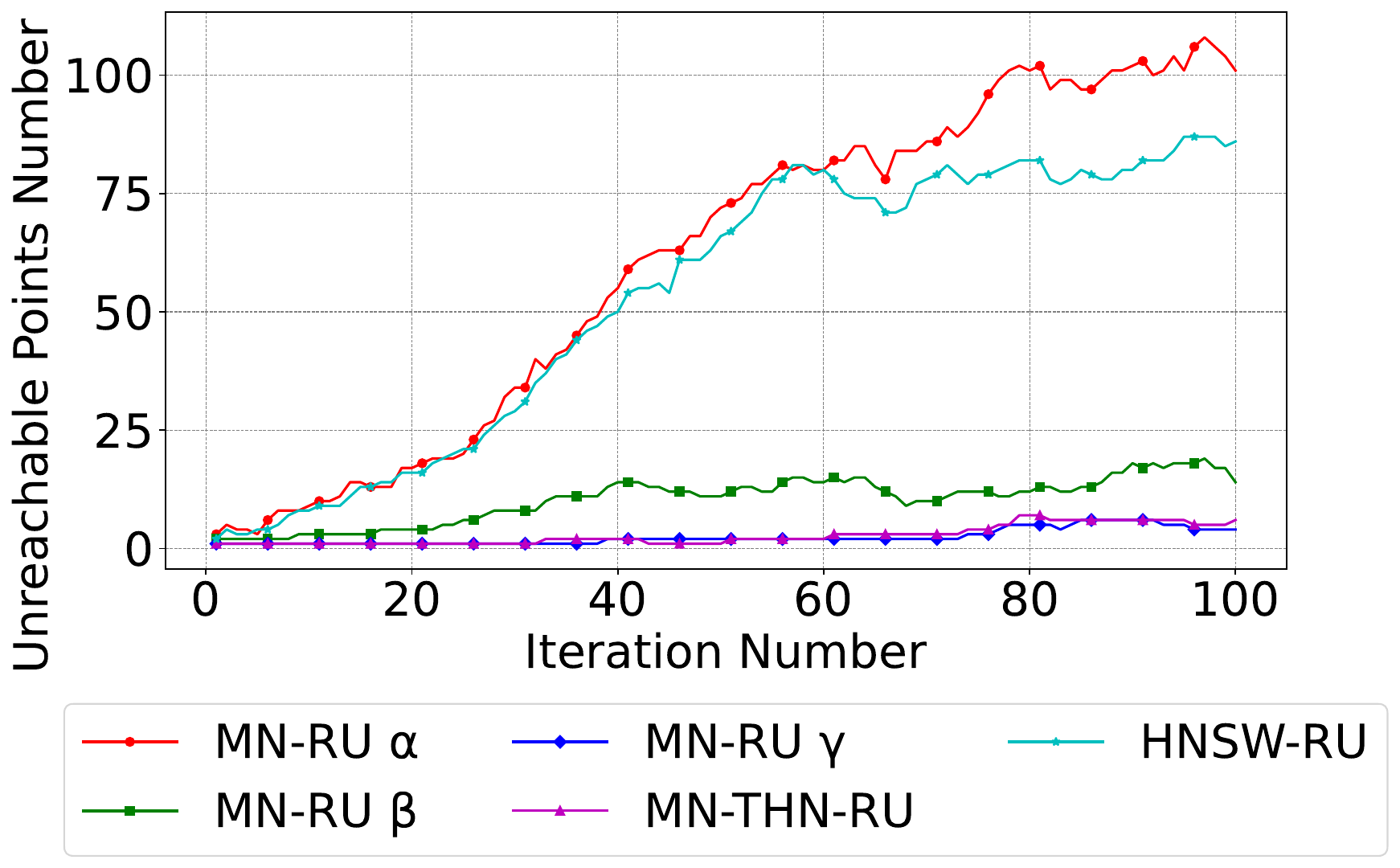}
        \caption*{(c) Sift}
    \end{minipage}
    \caption{Figure 8: Growth of unreachable points across various datasets in full\_coverage scenarios.}
    \label{fig:full_coverage_unreachable_points_growth}
    \vspace{-1em}
\end{figure*}

\begin{figure*}[ht]
    \centering
    \begin{minipage}[t]{0.32\textwidth}
        \centering
        \includegraphics[width=\textwidth]{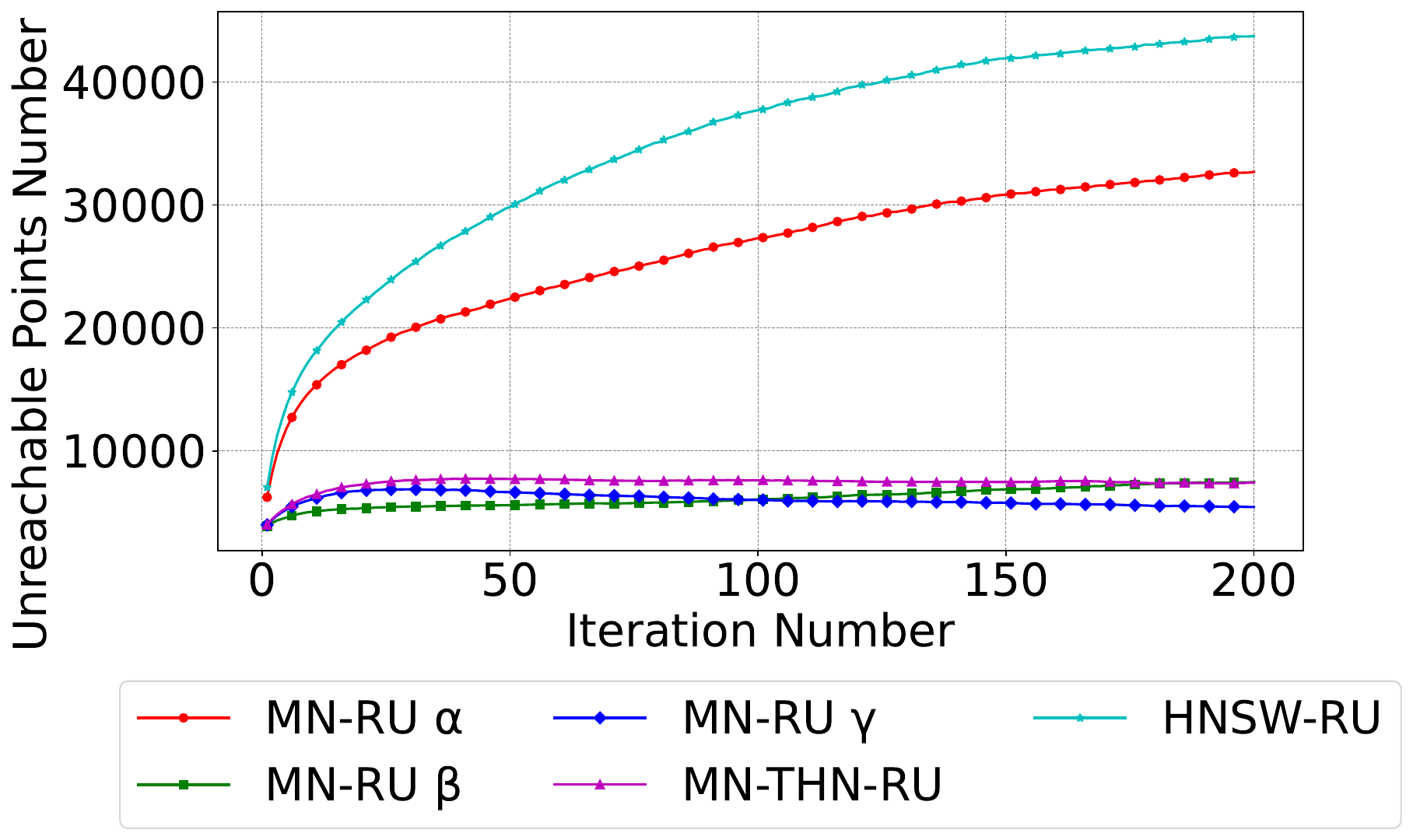}
        \caption*{(a) Gist}
    \end{minipage}
    \begin{minipage}[t]{0.32\textwidth}
        \centering
        \includegraphics[width=\textwidth]{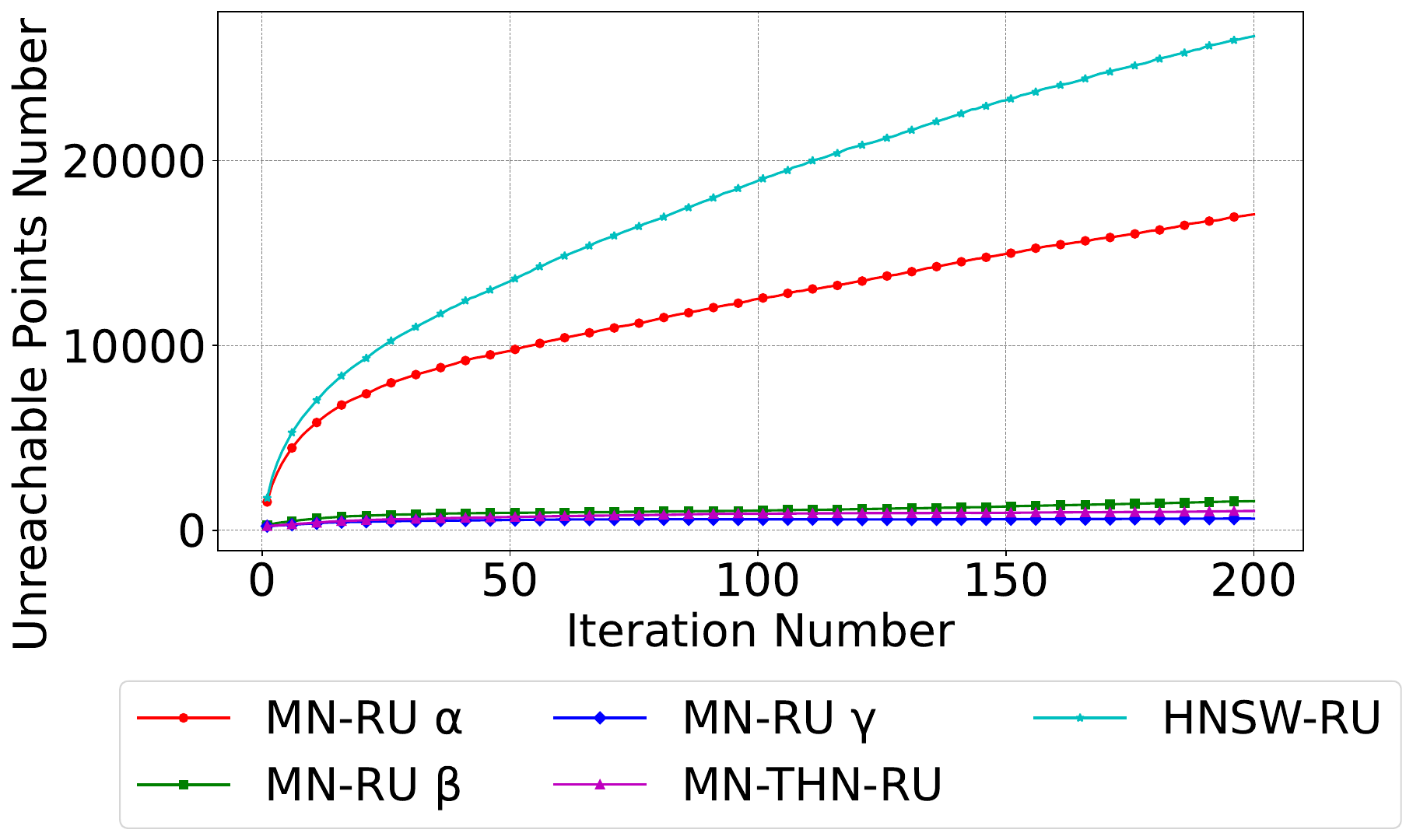}
        \caption*{(b) ImageNet}
    \end{minipage}
    \begin{minipage}[t]{0.32\textwidth}
        \centering
        \includegraphics[width=\textwidth]{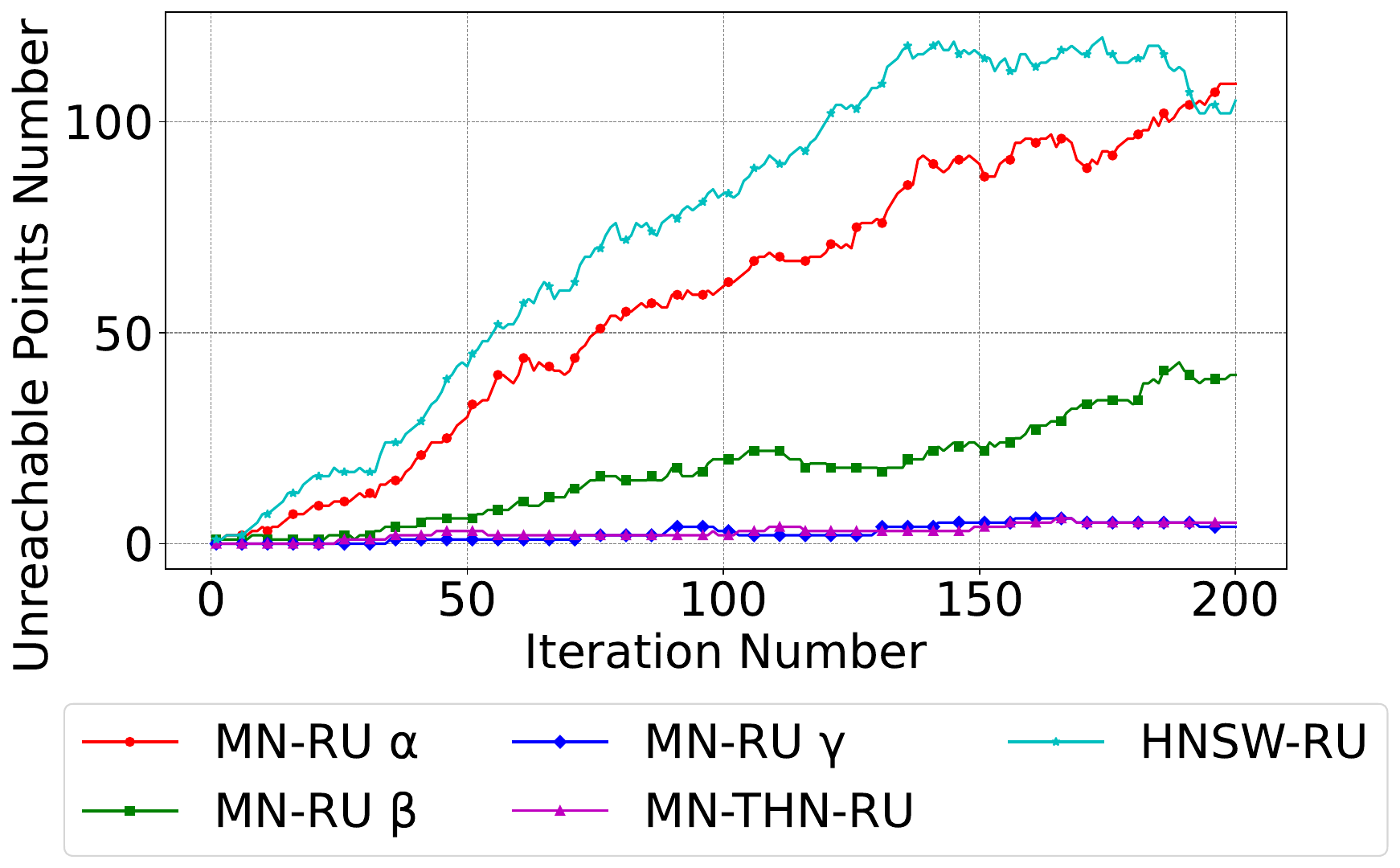}
        \caption*{(c) Sift}
    \end{minipage}
    \caption{Figure 9: Growth of unreachable points across various datasets in random scenarios.}
    \label{fig:random_unreachable_points_growth}
\vspace{-2em}
\end{figure*}


To understand the impact of unreachable points on search performance after updates, we conducted experiments on the Gist and ImageNet datasets. Figures \ref{fig:full_coverage_unreachable_points_growth} and \ref{fig:random_unreachable_points_growth} show that, after several updates, unreachable points can occupy a significant portion of these datasets, making the impact more observable.


Therefore, we investigated the search performance of indices built on Gist and ImageNet datasets under full\_coverage and random scenarios after updates to assess how unreachable points affect recall. The original Gist query set has only 1,000 queries and the ImageNet query set has 200, making it difficult to cover the entire dataset, especially for small values of $K$ ($K \leq 100$). This limits our ability to observe the impact of unreachable point growth on search accuracy.


To address this limitation, we adopted a more straightforward approach to test the impact of unreachable points on final recall: we used the entire original dataset as a query set. We performed nearest neighbor searches with $K=1$. Changing the parameter $ef\_$ to balance the recall and search time, we obtained the results shown in Figures \ref{fig:full_coverage_recall_gist_imageNet_endrecall} and \ref{fig:random_gist_imageNet_end_recall}. The results indicate that the MN-RU $\gamma$ and MN-THN-RU methods outperform HNSW-RU by achieving better accuracy within the same time frame. This demonstrates the efficacy of our methods in maintaining high precision and effectively reducing the number of unreachable points, leading to improved search accuracy, especially after many update operations.


\begin{figure*}
    \centering
    \begin{minipage}[t]{0.48\textwidth}
        \centering
        \subfloat[(a) Gist]{\includegraphics[width=0.5\textwidth]{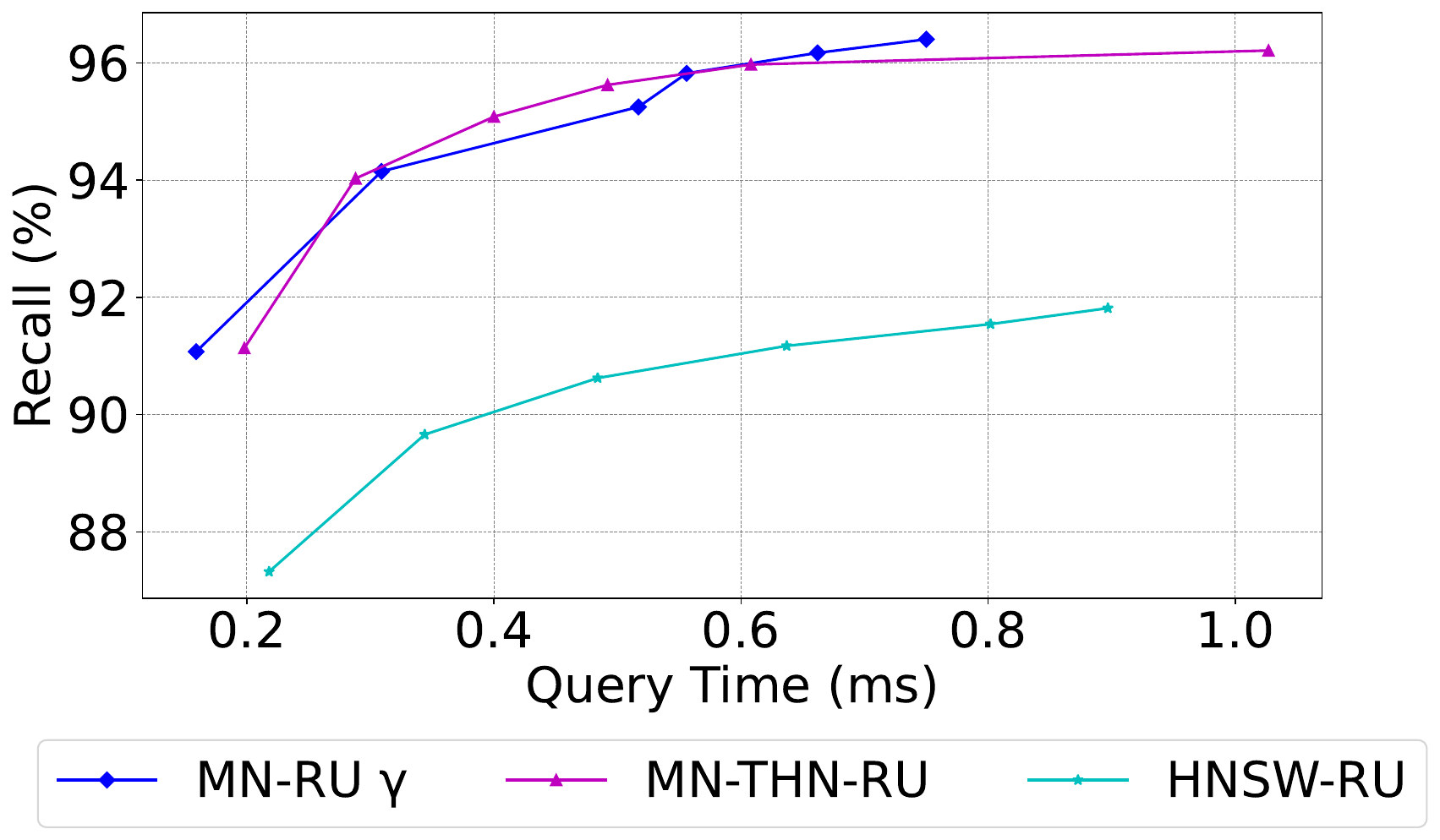}}
        \subfloat[(b) ImageNet]{\includegraphics[width=0.5\textwidth]{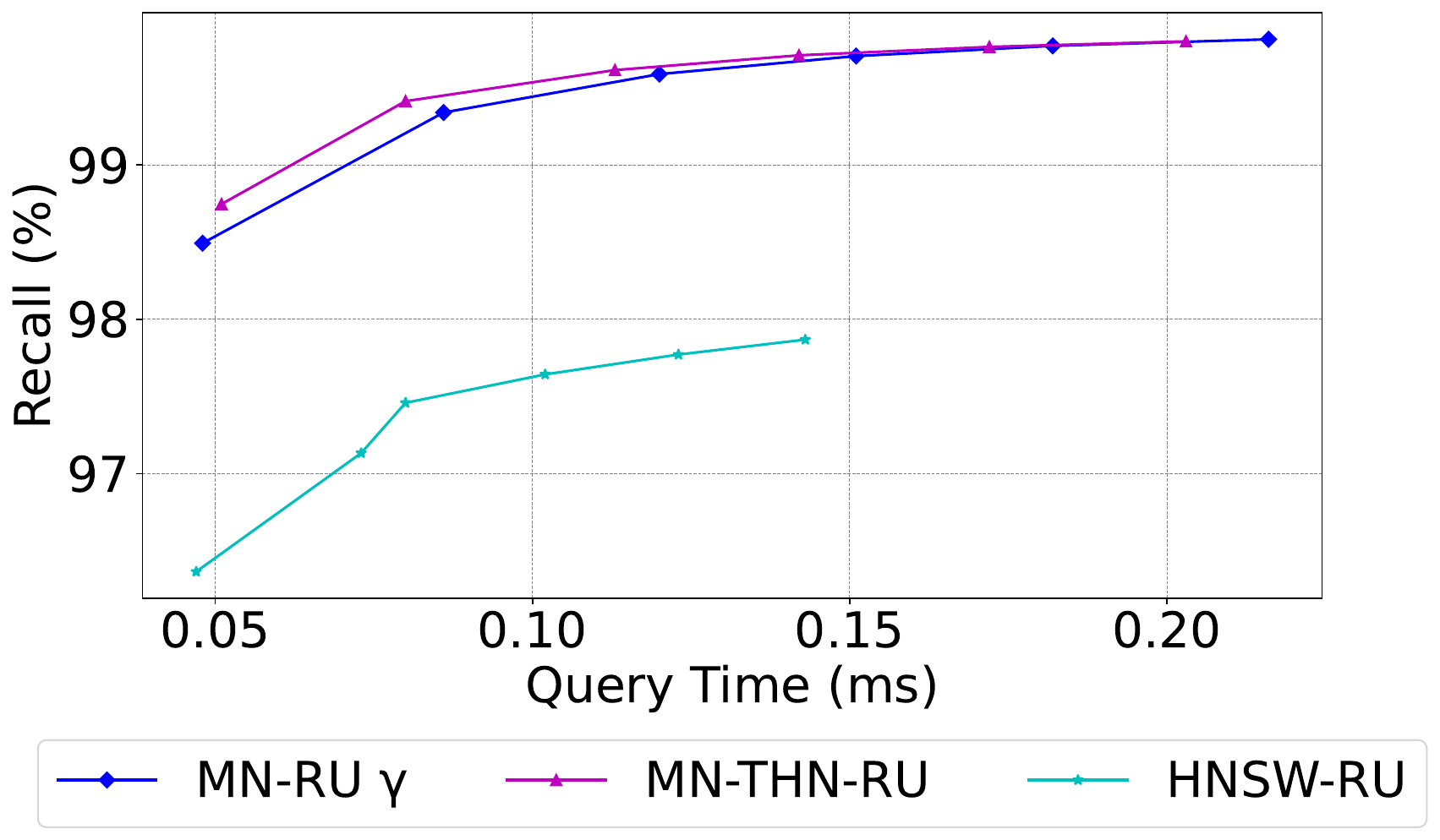}}
        \caption{Figure 10: Search performance following update operations in full\_coverage scenario using the Gist and ImageNet dataset.}
        \label{fig:full_coverage_recall_gist_imageNet_endrecall}
    \end{minipage}
    \begin{minipage}[t]{0.48\textwidth}
        \centering
        \subfloat[(a) Gist]{\includegraphics[width=0.5\textwidth]{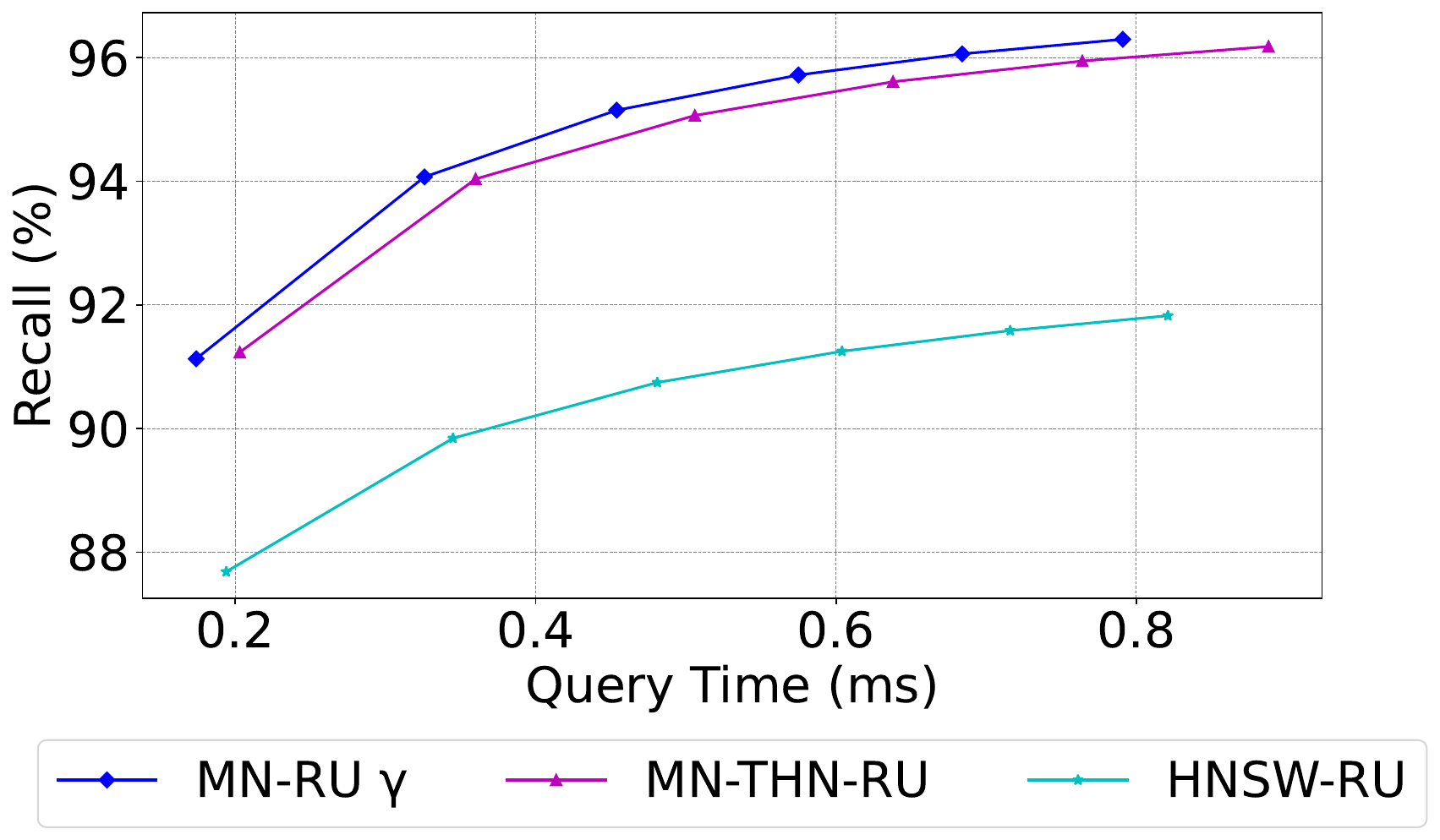}}
        \subfloat[(b) ImageNet]{\includegraphics[width=0.5\textwidth]{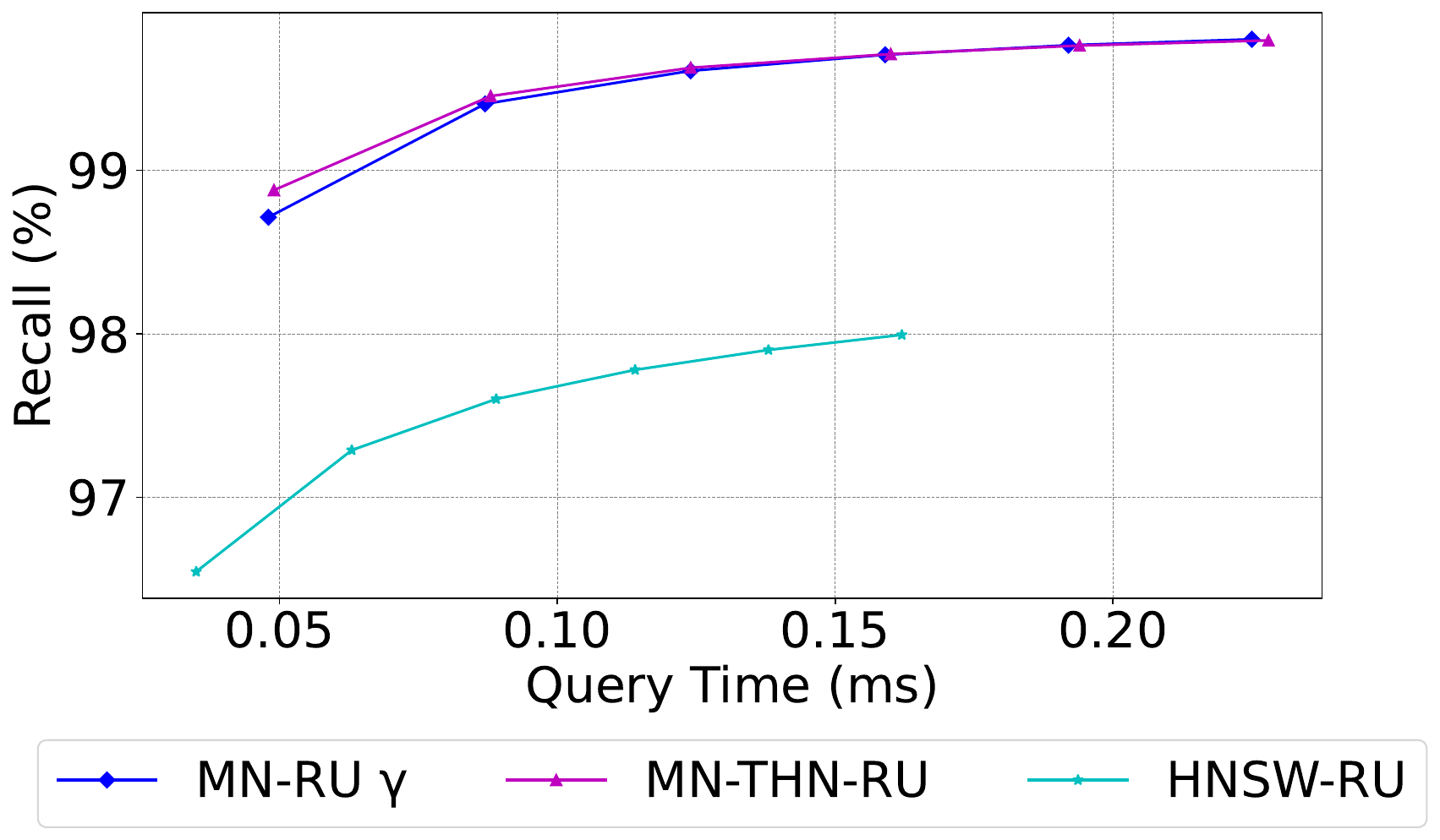}}
        \caption{Figure 11: Search performance following update operations in random scenario using the Gist and ImageNet dataset.}
        \label{fig:random_gist_imageNet_end_recall}
    \end{minipage}
    \vspace{-1em}
\end{figure*}


\begin{figure*}
    \centering
    \begin{minipage}[t]{0.33\textwidth}
        \centering
        \includegraphics[width=\textwidth]{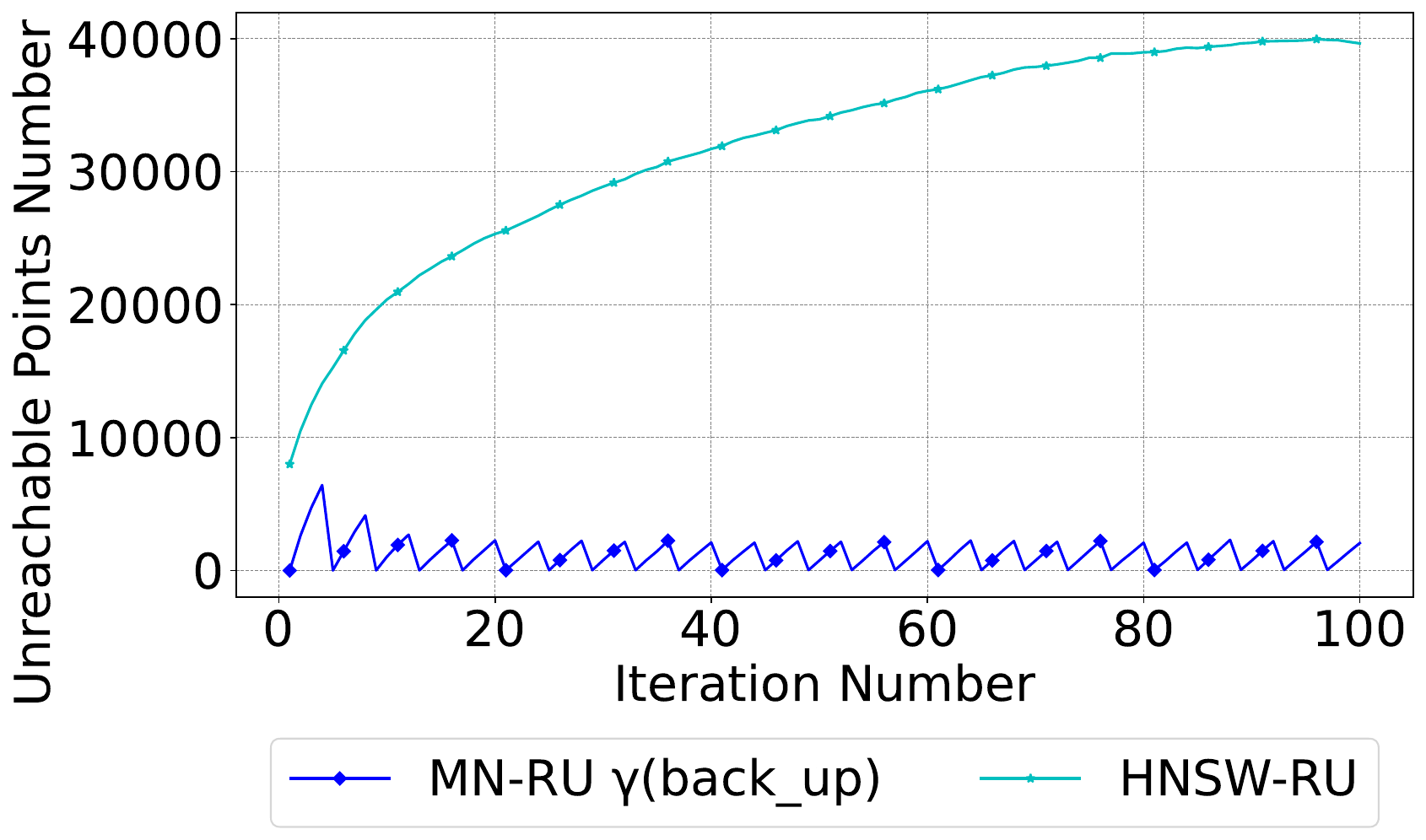}
        \caption{Figure 12: Growth of unreachable points between MN-RU $\gamma$ with back up index and the HNSW-RU in full\_coverage scenario using Gist dataset}
        \label{fig:back_up}
    \end{minipage}
    \begin{minipage}[t]{0.66\textwidth}
        \centering
        \subfloat[update time]{\includegraphics[width=0.48\textwidth,height=0.3\textwidth]{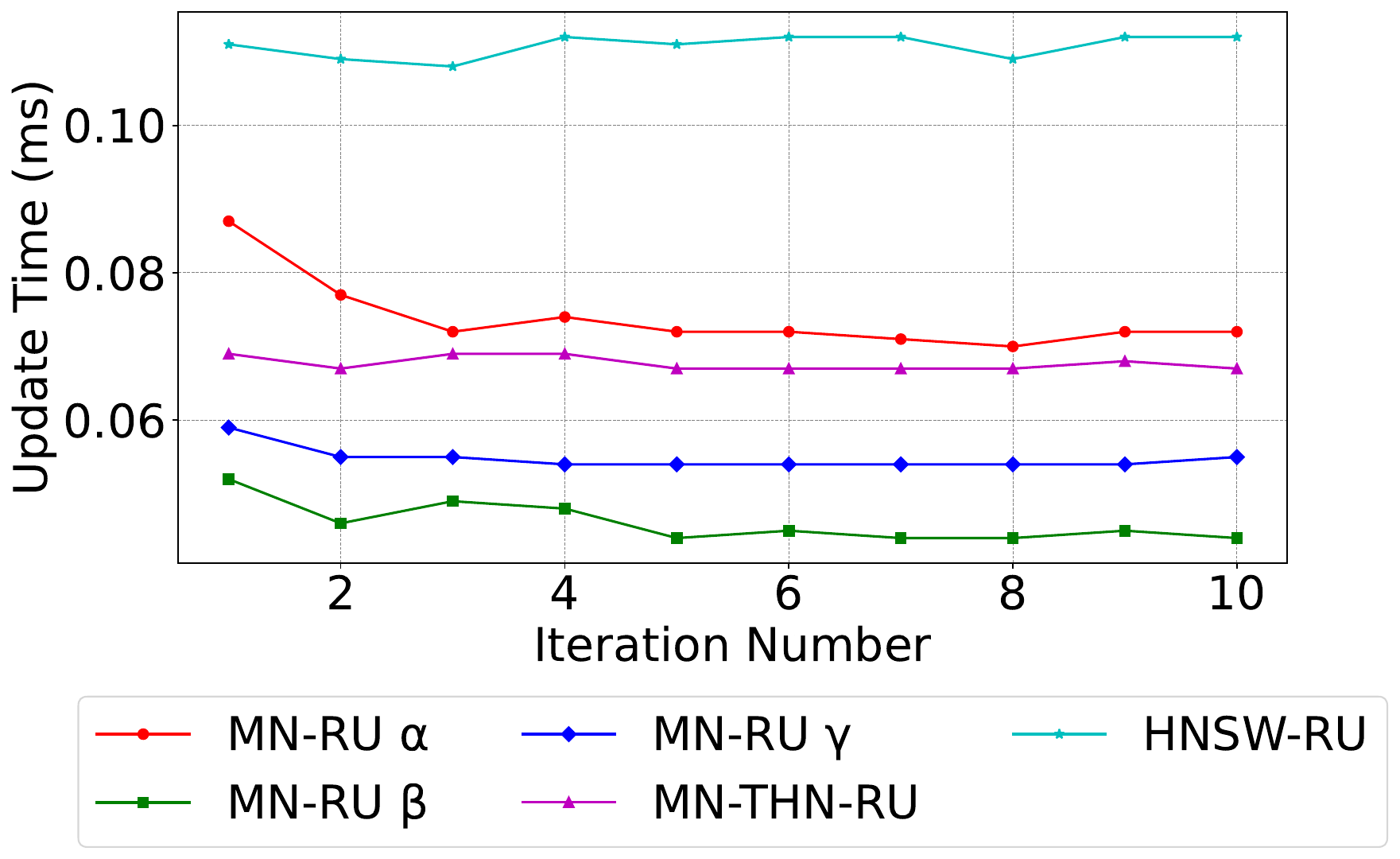}}
        \subfloat[unreachable points growth]{\includegraphics[width=0.5\textwidth,height=0.3\textwidth]{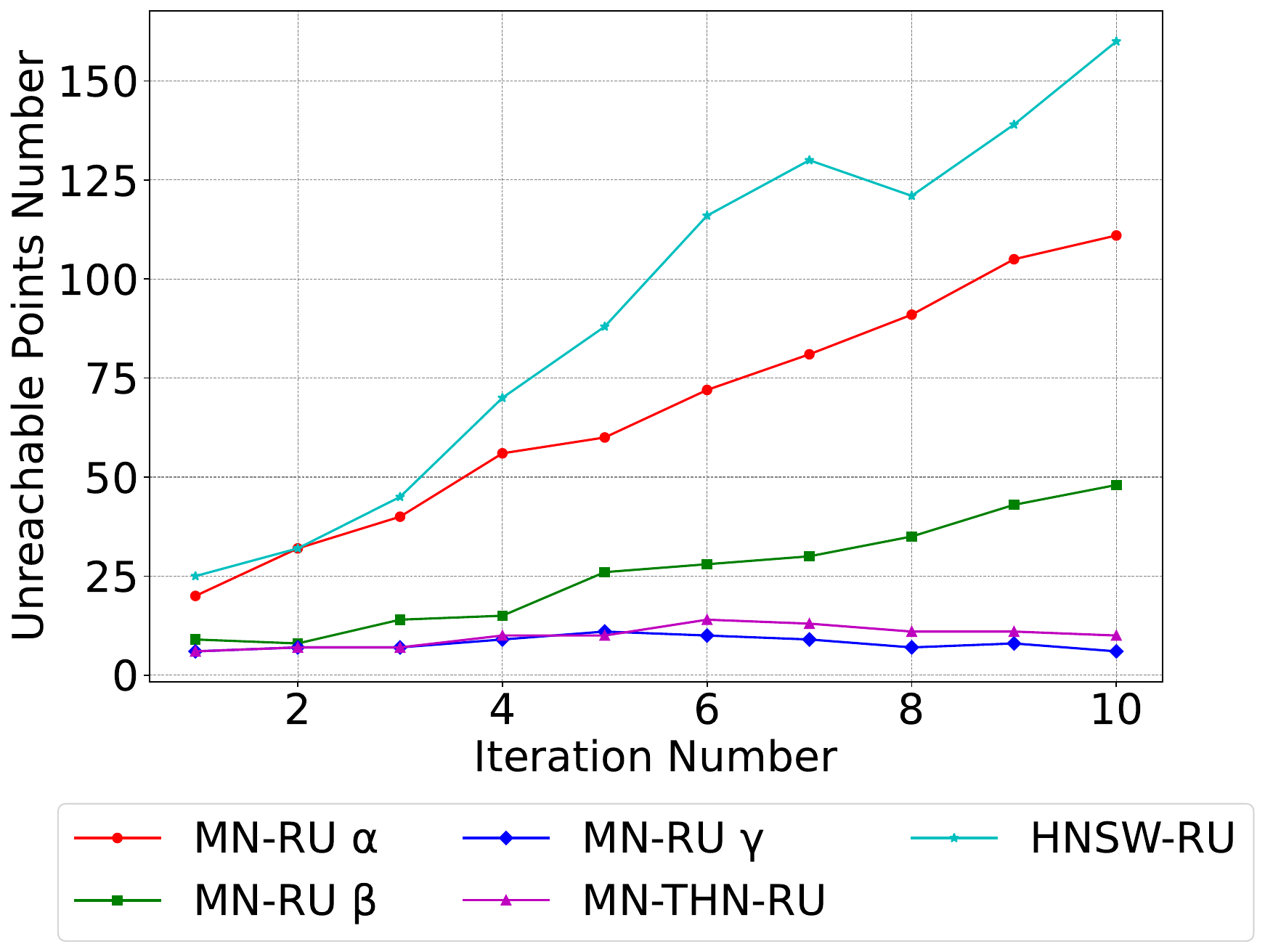}}
        \caption{Figure 13: Performance of update time and growth of unreachable points using the Sift\_2M dataset in new\_data scenario.}
        \label{fig:new_data_update_time_unreachable_points}
    \end{minipage}
\vspace{-2em}
\end{figure*}


To further validate the effectiveness of our backup index construction methods and the algorithm \ref{alg:dual_search} illustrated in Figure \ref{fig:architecture}, we experimented on the GIST dataset under the full\_coverage scenario. This experiment compared the growth of unreachable points between the MN-RU $\gamma$ with the backup index method and the original HNSW-RU. We set the parameter $\tau$ to 40,000, which means that the backup index was rebuilt after every four iterations. As shown in Figure \ref{fig:back_up}, the MN-RU $\gamma$ with the backup index method significantly reduced the number of unreachable points compared to the original HNSW-RU. Therefore, our method with the backup index outperforms the original HNSW-RU baseline.

\section{Conclusion}
This paper addresses HNSW replaced\_update limitations in real-time deletions and insertions, leading to the `unreachable points phenomenon' and reduced efficiency. Our proposed MN-THN-RU and MN-RU $\gamma$ algorithms with the backup index method effectively mitigate these issues by enhancing the efficiency of mixed operations and maintaining better graph connectivity.
Extensive experimental validation demonstrates that our methods outperform the native HNSW strategy, significantly reducing the number of unreachable points and improving the update speed across various datasets. These improvements make our approach highly practical for real-world applications requiring dynamic real-time updates and high search accuracy.


\bibliographystyle{IEEEtran}
\bibliography{acml24}

\end{document}